\documentclass[letterpaper,twocolumn,10pt]{article}
\usepackage{usenix-2020-09}

\usepackage{tikz}
\usepackage{amsmath}

\usepackage{times,url,color,soul,xspace,enumitem}
\usepackage{graphicx}
\usepackage{caption}
\usepackage{subcaption}
\usepackage{comment}
\usepackage{xspace}
\usepackage{tikz}
\usepackage{amsmath}
\usepackage[inline,draft,nomargin,index]{fixme}
\usepackage{xurl}

\usepackage{listings}
\usepackage{cleveref}
\crefformat{section}{\S#2#1#3} 
\crefformat{subsection}{\S#2#1#3}
\crefformat{subsubsection}{\S#2#1#3}

\FXRegisterAuthor{giu}{agiu}{\textcolor{red}{Giu}}

\FXRegisterAuthor{atr}{aatr}{\textcolor{blue}{Atr}}

\FXRegisterAuthor{lin}{alin}{\textcolor{green}{Lin}}

\FXRegisterAuthor{nty}{anty}{\textcolor{orange}{Nty}}

\begin{document}

\date{}


\title{\Large \bf Understanding NVMe Zoned Namespace (ZNS) Flash SSD Storage Devices}
\author{
{\rm Nick Tehrany}\\
Vrije Universiteit Amsterdam\\
Delft University of Technology
\and
{\rm Animesh Trivedi}\\
Vrije Universiteit Amsterdam
}

\maketitle

\begin{abstract}
    The standardization of NVMe Zoned Namespaces (ZNS) in the NVMe 2.0 specification presents a unique new addition to storage devices. Unlike traditional SSDs, where the flash media management idiosyncrasies are hidden behind a flash translation layer (FTL) inside the device, ZNS devices push certain operations regarding data placement and garbage collection out from the device to the host. This allows the host to achieve more optimal data placement and predictable garbage collection overheads, along with lower device write amplification. Thus, additionally increasing flash media lifetime. As a result, ZNS devices are gaining significant attention in the research community.
    
However, with the current software stack there are numerous ways of integrating ZNS devices into a host system. In this work, we begin to systematically analyze the integration options, report on the current software support for ZNS devices in the Linux Kernel, and provide an initial set of performance measurements. Our main findings show that larger I/O sizes are required to saturate the ZNS device bandwidth, and configuration of the I/O scheduler can provide workload dependent performance gains, requiring careful consideration of ZNS integration and configuration depending on the application-workload and its access patterns.
Our dataset and code are aavailable at \url{https://github.com/nicktehrany/ZNS-Study}. 
\end{abstract}

\section{Introduction}

The introduction of flash storage provided significant changes in the storage hierarchy. Achieving as low as single
digit $\mu$-second latency, several GB/s of bandwidth, and millions of I/O operations per second
(IOPS)~\cite{2022-samsung-zand,2022-intel-p}, they offer significant performance gains over prior storage technologies,
such as Hard Disk Drives (HDDs). Flash storage is organized in pages (typically 16KiB in
size)~\cite{2008-Agrawal-Design-Tradeoff-SSD}, representing the unit of read and write accesses, of which multiple pages
are combined into a block (typically multiple MiB in size). Blocks are further packed in planes and dies to manage data
and control connectivity to the host. Flash pages do not support in-place updates. As a result, pages have to be erased
prior to being written again. However, erase operations require substantially more time than read and write
operations~\cite{2009-Gupta-DFTL}. Therefore, erase operations are done at block granularity to amortize the erase
overhead. Additionally, flash storage requires pages within a block to be written sequentially. 

Flash storage therefore includes complex firmware, called the Flash Translation Layer (FTL), to provide the seemingly
in-place updates of data and hide the sequential write constraints of devices by exposing a sector/page-addressable
SSD~\cite{2008-Agrawal-Design-Tradeoff-SSD}. Furthermore, a crucial task of the FTL is to run 
\textit{garbage collection} (GC), in order to erase blocks with invalid pages and free up space. For this, the FTL reads out valid pages of
data from a block and relocates the data to a free block, followed by erasing of the original block. Garbage collection
is triggered periodically, or when the device is running low on free space to write data.

The resulting interface exposed by conventional flash storage allows it to mimic the behavior of HDDs, thus requiring
no changes in the host storage software to access the underlying flash storage. However, recent research results made evident that hiding the flash management complexities
from the host leads to suboptimal data placement, unpredictable performance overheads, and shortens the lifetime of
flash devices~\cite{2017-He-SSD-Unwritten-Contract}. Therefore, researchers have proposed to open the flash storage
interface and expose device internals to the host. This allows for the host to optimize storage management with
workload-specific decisions~\cite{2017-fast-lightnvm,2012-asplos-moneta,2014-hotstorage-multistream-ssd,2009-fast-dfs}.

Zoned Namespace (ZNS) SSDs are the latest addition in these efforts, which are now standardized in the NVMe 2.0 specification~\cite{2022-nvme-spec} and are commercially available~\cite{2021-Bjorling-ZNS}. In order to better match the underlying properties of flash chips, ZNS exposes the address space with numerous \textit{zones}, where each zone requires append-only sequential writes. Zones are aligned to the erase unit of a block. With this new interface, the host storage software is now responsible for resetting a zone (i.e., trigger garbage collection) after which a zone becomes writable again from the starting address. Apart from the write restrictions, there are operational parameters such as the zone capacity, the maximum number of active zones limits, and the append limits, which the host software must be aware of.

\begin{figure}[!t] 
    \centering
    \includegraphics[width=0.49\textwidth]{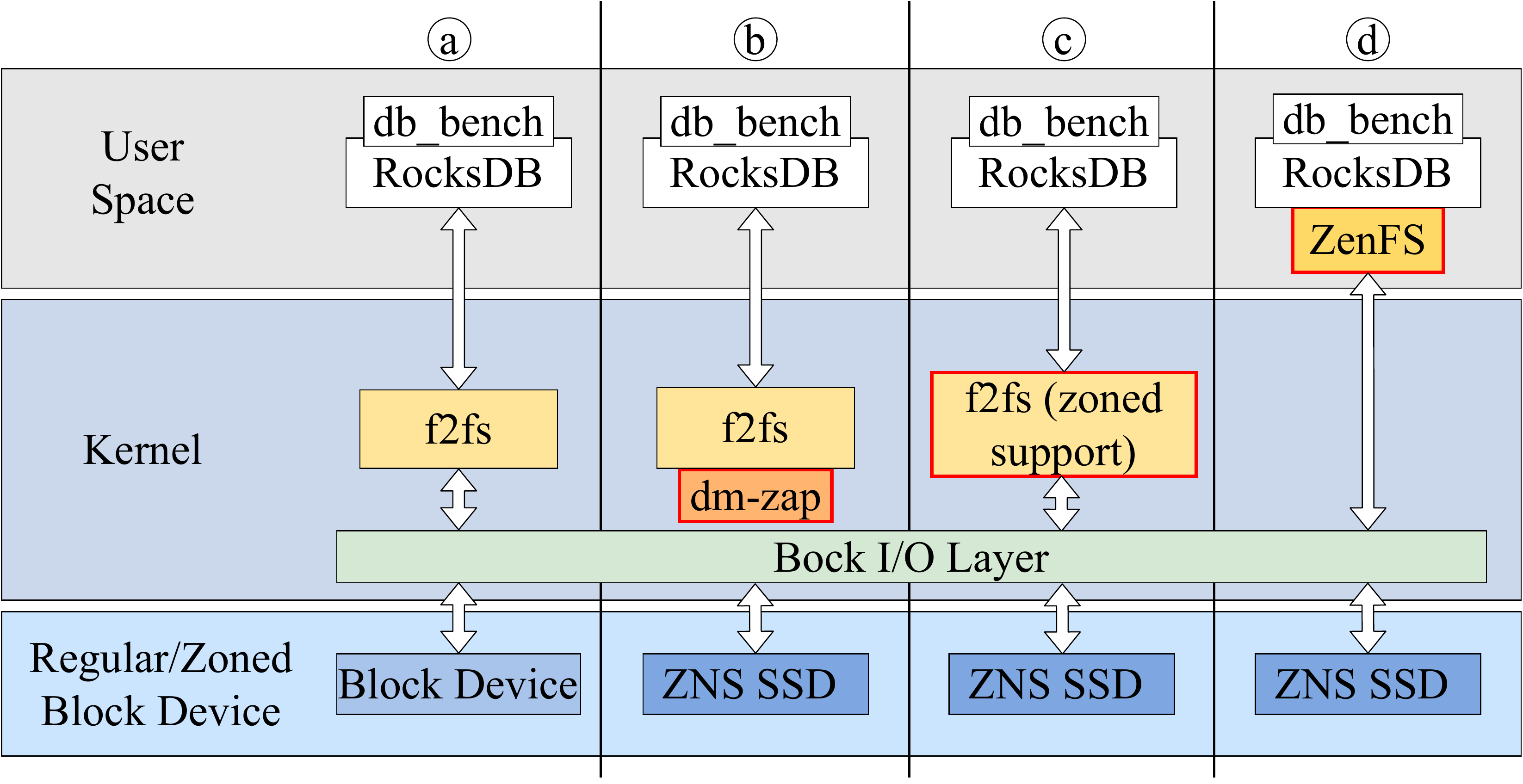}
    \caption{Integration of conventional block devices into the host software stack (a) compared to the three levels of integration for ZNS SSDs (b)-(d) into the software stack.}
    \label{fig:ZNS_configurations}
\end{figure}

The new interface of these devices necessitates changes to the host storage stack. However, there is more than one way these devices can be integrated within systems. In a classical setup as shown in Figure~\ref{fig:ZNS_configurations}(a), at the bottom is a block device with a file system on top of it, and an application running on top of the file system. Integrating ZNS devices into this storage stack can be done in three different configurations. Firstly, integration at the block-device level (Figure~\ref{fig:ZNS_configurations}(b)), while keeping the rest of the stack above the same. An example of this setup is the \textit{dm-zap} zoned device mapper project within the Linux Kernel~\cite{2022-dmzap-git}. This way of integration is the least intrusive one and requires the minimum amount of changes to anything running on top of the block device. Secondly, integration at the file system level (Figure~\ref{fig:ZNS_configurations}(c)). For this integration a file system is made aware of the zoned device characteristics such as zone capacity, number of active zones limits, and append limits. Example of such a project is the added ZNS support in f2fs~\cite{2021-Bjorling-ZNS}. With such integration the knowledge and required changes for ZNS devices are pushed higher in the stack, from block-device level to the file system level. Lastly, ZNS-aware application integration (Figure~\ref{fig:ZNS_configurations}(d)). In this case, there are ZNS-specific application-level changes at the very top of the storage stack. By pushing the customization higher up in the stack, the expectation is to deliver better performance together with the best case application-specific customization and integration with ZNS devices. An example of such an integration is the ZenFS file system module for RocksDB and MySQL~\cite{2021-Bjorling-ZNS,2022-zenfs-git}. 

With such configuration possibilities, it is not immediately clear which integration one should choose for their workload. In this research work, we aim to systematically understand the impact of the ZNS integration in the host storage system. This work is largely inspired by related work that has provided unwritten contracts of storage devices for Optane based SSDs~\cite{2019-Wu-Unwritten_Contract_Optane} and flash based SSDs~\cite{2017-He-SSD-Unwritten-Contract} with guidelines for developers to optimize storage performance. Before we can synthesize actionable design guidelines for storage stack developers, in this work, we first start with systematic benchmarking in the presence of the OS I/O scheduler. In particular, we make the following contributions:

\begin{itemize}
    \item We provide information on the newly standardized NVMe ZNS devices, how these devices work, how they are integrated into the host storage stack, and how existing applications are modified to support ZNS devices.
    \item We measure the block-level ZNS device performance, comparing it to conventional block devices in terms of achievable IOPs and bandwidth, and benchmark the possible scheduler configurations for ZNS devices, depicting their implications and limitations. 
    \item We present pitfalls and failed experiments during this evaluation in an effort for others to learn from and to avoid the obstacles we encountered.
    \item We provide a set of initial guidelines for optimizing ZNS integration into systems, and propose several future work ideas to further explore ZNS integration implications and expand our initial set of guidelines.
    \item All collected datasets and benchmarking scripts are made publicly available at \url{https://github.com/nicktehrany/ZNS-Study}. The appendix provides more detailed setup and benchmarking information. 
\end{itemize}

The remainder of the paper is organized as follows. Section~\ref{sec:background} provides background information on ZNS devices, and their integration into systems. Next, Sections~\ref{sec:exp_setup} explains the experimental setup, followed by the first round of unsuccessful experiments in Section~\ref{sec:unsuccessful_exps}. Based on this we provide an adapted experimental setup in Section~\ref{sec:adapted_setup}, and Section~\ref{sec:evaluation} presents the various benchmarks. Lastly, Section~\ref{sec:related} provides the related work, followed by future work ideas in Section~\ref{sec:future_work}, and Section~\ref{sec:conclusion} concludes the paper.

\section{Background}\label{sec:background}
Applications and file systems rely on the Linux block layer to provide interfaces and abstractions for accessing the underlying storage media. Originally designed to match the hardware characteristics of HDDs the block layer presents the storage as a linear address space, allowing for sequential and random writes. Flash storage however has different write constraints due to its architecture. Relying on the FTL to hide the device management idiosyncrasies however leads to negative performance impacts due to the unpredictable performance from the device garbage collection~\cite{2015-Kim-SLO_complying_ssds,2014-yang-dont_stack_log_on_log}, large tail latency it causes~\cite{2013-Dean-tail_at_scale}, and increased write amplification~\cite{2014-Desnoyers-Analytic_Models_SSD}. The increased write amplification additionally reduces the device lifetime, since flash cells on SSDs have limited program/erase cycles.

Furthermore, garbage collection requires the device to maintain a certain amount of free space, called the \textit{overprovisioning space}, such that the FTL is able to move valid pages. Most commonly used overprovisioning takes between 10-28\% of the device capacity. One of the possible FTL design uses a fully-associative mapping of host logical block addresses (LBAs) to physical addresses~\cite{2009-Gupta-DFTL} in order to provide the LBAs of the page(s) that contain valid data. Such a design requires significant resources in order to store all mappings. 

\subsection{Zoned Storage}\label{sec:zoned_storage}
The arrival of ZNS SSDs eliminates the need for on device garbage collection done by the FTL, pushing this responsibility to the host. This provides the host with more opportunity for optimized data placement, through mechanisms such as data grouping, and makes garbage collection overheads predictable. The concept of exposing storage as zones is not new, as it was already introduced when Shingled Magnetic Recording (SMR) HDDs~\cite{Feldman2013ShingledMR,Gibson2011PrinciplesOO,Suresh2012ShingledMR} appeared, which also enforce a sequential write constraint. The zoned storage model was established with the addition of Zoned Block Device (ZBD) support in the Linux Kernel 4.10.0~\cite{2021-ZNS-ZBD_docs}, in an effort to avoid the mismatch between the block layer and the sequential write constraint on devices. 

The standards defining the management of SMR HDDs in the zoned storage model came through the Zoned Device ATA Command Set (ZAC)~\cite{2015-ZAC} and the Zoned Block Command (ZBC)~\cite{2014-ZBC} specifications. Since zones require sequential writing, the address of the current write is managed with a \textit{write pointer}. The write pointer is incremented to the LBA of the next write only after a successful write. In order to manage zones, each zone has a state associated to it. These are to identify the condition of a zone, which can be any of the following; \textit{EMPTY} to indicate a zone is empty, \textit{OPEN} which is required for a zone to allow writes, \textit{FULL} to indicate the write pointer is at the zone capacity, \textit{CLOSED} which is used to release resources for the zone (e.g., write buffers on the ZNS device) without resetting a zone, this additionally does not allow writes to continue in the zone until it is transitioned to \textit{OPEN} again, \textit{OFFLINE} which makes all data in the zone inaccessible until the zone is reset, and \textit{READ-ONLY}. The command sets provide the proper mechanisms to transition zones between any of the states.

While the majority of available space for both SMR and ZNS devices is utilized by sequential write required zones, they can also expose a limited amount of randomly writable area. This is mainly intended for metadata as this space only occupies a very small percentage of the device capacity. For example, the ZNS device used in this evaluation could expose 4 zones as randomly writable, equivalent to approximately 4GiB, compared to the total device size of 7.2TiB.

\begin{figure}[]
    \centering
    \includegraphics[width=0.48\textwidth]{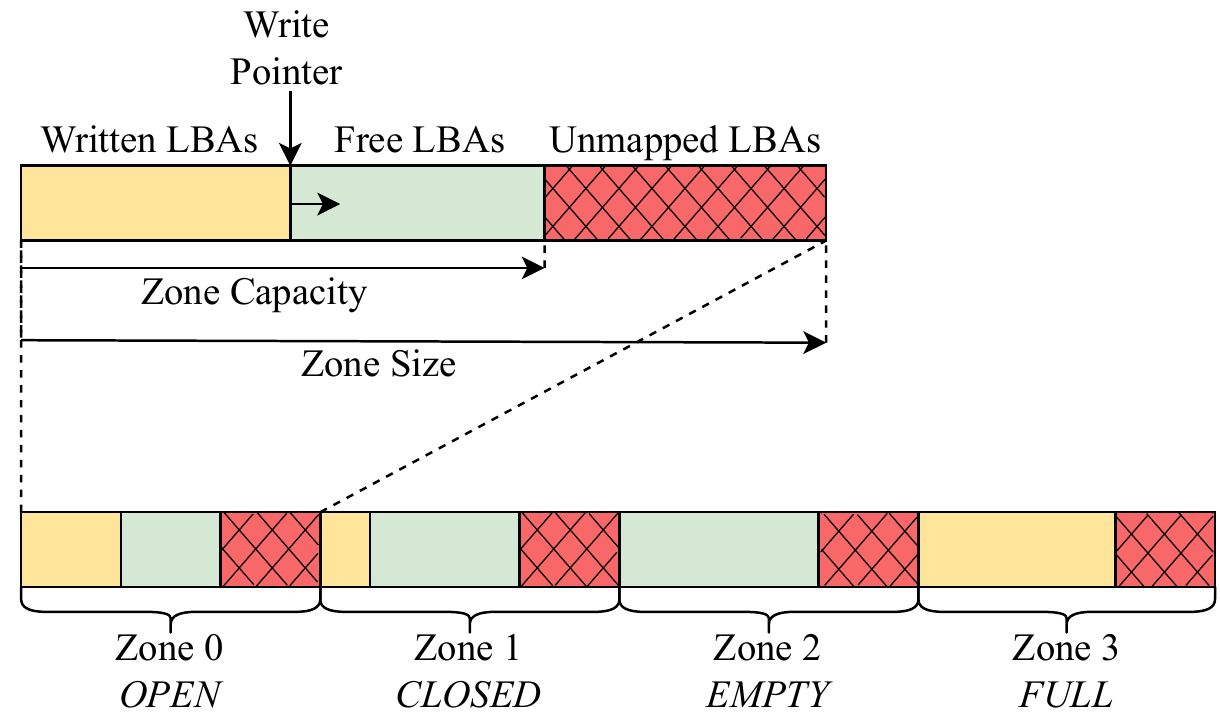}
    \caption{Layout of a ZNS SSD, depicting zone capacity, write pointer, and zone states associated to each zone. Adapted from~\cite{2021-Bjorling-ZNS}.}
    \label{fig:ZNS_HW}
\end{figure}

The newly introduced ZNS SSDs are standardized through the NVMe specification~\cite{2022-nvme-spec}, which builds on the foundations established with ZBD support. While the zoned storage model aims to provide a unified software stack for all zoned devices (SMR and ZNS devices), the NVMe specification introduces several new concepts particular to ZNS devices. Firstly, it defines a \textit{zone capacity} for each zone, stating the usable capacity of a zone, which is less than or equal to the size of the zone. Figure~\ref{fig:ZNS_HW} shows an example layout of zones on a ZNS SSD and each zone's associated state. Zone capacity is kept separate from the zone size such that the capacity can be aligned to the erase unit of the underlying flash media, and such that the zone size can be kept as a power of two value. This is required for easier conversions between LBAs and zone offsets.

Secondly, the \textit{active zones} limit specifies the maximum number of zones that can be active (in \textit{OPEN} or
\textit{CLOSED} state) at any point in time. As the SSD requires to allocate resources for each active zone, such as the
write buffers, it enforces a maximum on the active zones. Lastly, the NVMe specification introduces the \textit{zone
append} command~\cite{bjorling2020zoneappend}, providing the option for the host to maintain multiple outstanding I/Os
in a zone. The ZNS controller writes the data at the write pointer, and returns the LBA of the write to the host.
Therefore, making append especially beneficial if a large number of small writes are issued. 

If applications do not utilize the append command, they are required to ensure correct ordering among I/Os such that writes happen at the correct LBA, equal to the write pointer of the zone. ZNS devices additionally require the use of direct I/O, bypassing the page cache. This enforces to align with the sequential write constraint of zones, such that pages from the page cache are not written at out-of-order LBAs.

\subsection{ZNS I/O Scheduling}
Adhering to the sequential write requirement with ZNS devices under multiple outstanding I/O requests requires an appropriate scheduler to be set. It is responsible for ensuring writes are submitted at consecutive LBAs on the device~\cite{2021-ZNS-documentation}. However, requests can additionally be reordered on the device~\cite{2022-nvme-spec}, making the host responsible for ensuring command ordering. For this the \textit{mq-deadline} scheduler has to be enabled. It enforces that just a single write I/O is submitted and holds all subsequent I/Os until completion of the submitted I/O. This allows to submit numerous asynchronous write requests while adhering to the sequential write requirement and enforcing correct command ordering.

Additionally, the scheduler set to \textit{none} can also be used with ZNS devices, bypassing the Linux Kernel I/O scheduler, however this does not enforce sequential write ordering or command ordering, as I/O requests are directly issued to the device. Hence, if this scheduler is set writes have to be issued synchronously and at the correct LBA. As ZNS devices only enforce sequential write ordering, reading can be done with both schedulers asynchronously under any number of outstanding I/Os, and random or sequential accesses.

\subsection{ZNS Application Integration}
With the zoned storage model providing a unified software stack for SMR and ZNS devices and support for SMR having been in numerous applications for some time, required changes for added support of ZNS devices was minimal.  We focus primarily on f2fs (with f2fs-tools~\footnote{f2fs-tools provides the mkfs.f2fs functionality to format the storage device in order to mount the f2fs file system. ZNS support was added in version 1.14.0. Available at \url{https://git.kernel.org/pub/scm/linux/kernel/git/jaegeuk/f2fs-tools.git/about}})~\cite{2015-Changman-f2fs}, and ZenFS (commit \textit{5ca1df7})~\cite{2021-Bjorling-ZNS}, a RocksDB storage backend for zoned storage devices.

\noindent \textbf{f2fs:} f2fs had existing support for the ZBC/ZAC specification~\cite{2016-Axboe-f2fs_ZBC_patch}, making the changes for supporting ZNS devices minimal~\cite{2021-Bjorling-ZNS}. The changes include adding the zone capacity and limiting the maximum number of active zones. f2fs manages the storage as a number of segments (typically 2MiB), of which one or more are contained in a section. Sections are the erase unit of f2fs, and segments in a section are written sequentially. The segments are aligned to the zone size, such that they do not span across zones. Since ZNS devices have a zone capacity, which is possibly smaller than the zone size, an additional segment type \textit{unusable} was added in order identify segments outside the usable zone capacity. Partially usable segments are also supported with the \textit{partial} segment type, in order to fully utilize the entire zone capacity if a segment is not aligned to zone capacity. 

The maximum active zones was already implemented by limiting the maximum number of open segments at any point in time. By default, this is set to 6, however if the device supports less active zones, f2fs will decrease this at file system creation time. While f2fs supports ZNS devices, it requires an additional randomly writable block device for metadata, which is updated in-place, as well as caching of writes prior to writing to the zoned device.

\noindent \textbf{ZenFS:} ZenFS provides the file system plugin for RocksDB to utilize zoned block devices. RocksDB is an LSM-tree based persistent key-value store~\cite{2017-Dong-Optimizing-RocksDB,2022-Rocksdb-src} optimized for flash based SSDs. It works by maintaining tables at different levels in the LSM tree, of which the first level is in memory and all other levels are on the storage device. Writes initially go into the table in the first level, called the \textit{memtable}, which gets flushed to the next level periodically or when it is full. Flushing will merge the flushed table with one from the next level, such that keys are ordered and do not overlap. This process is called \textit{compaction}. Tables at lower levels than the memtable are referred to as Sorted String Tables (SSTs). SSTs are immutable, written sequentially, and erased as a single unit, hence making it a flash friendly architecture~\cite{2021-Bjorling-ZNS}.

With zoned storage devices the RocksDB data files need to be aligned to the zone capacity for most efficient device utilization. ZenFS maps RocksDB data files to a number of \textit{extents}, which are contiguous regions that are written sequentially. Extents are written to a single zone, such that they do not span across multiple zones, and multiple extents can be written into one zone, depending on the extent size and zone capacity. Selection of extent placement into zones relies on the provided lifetime hints that RocksDB gives with its data files. ZenFS places an extent into a zone where the to be written extent's lifetime is smaller than the largest lifetime of the other extents in the zone, such that it is not unnecessarily delaying the resetting of a zone. ZenFS resets a zone when all the files that have extents in that particular zone have been invalidated.

While RocksDB provides the option to set the maximum size for data files, data files will not have precisely this size due to compaction resulting in varying sized data files. ZenFS manages this by setting a configurable utilization percentage for the zones, which it fills up to this percentage, leaving space if files are larger than specified. While ZenFS requires at least 3 zones to run, of which one is for journaling, another is the metadata, and the last is a data zone, if the device supports more active zones, the active zones can be increased by setting a larger number of concurrent compactions in RocksDB. This can be up to the value of maximum active zones (minus the metadata and journaling zone), however performance gains for more than 12 active zones on particular ZNS devices are insignificant~\cite{2021-Bjorling-ZNS}.

\section{Experimental Setup}\label{sec:exp_setup}
\begin{table*}[]
    \centering
    \begin{tabular}{||c|c|c|c||}
        \hline
        & Optane SSD & Samsung SSD & ZNS SSD \\ \hline \hline
        Model & INTEL SSDPE21D280GA & Samsung SSD 980 PRO & WZS4C8T4TDSP303 \\
        Media Size & 260.8GiB & 1.8TiB & 7.2TiB \\
        Usable Capacity & 260.8GiB & 1.8TiB & 3.8TiB \\
        Sector Size & 512B & 512B & 512B \\
        Zone Size & - & - & 2048MiB \\
        Zone Capacity & - & - & 1077MiB \\
        Number of Zones & - & - & 3688 \\
        \hline
    \end{tabular}
    \caption{SSD architecture of the three utilized devices during experimentation. ZNS information depicts the zoned namespace on the ZNS device.}
    \vspace{-0.4cm}
    \label{tab:SSD_architecture}
\end{table*}

For the experiments we utilize a ZNS device, whose details and properties are depicted in Table~\ref{tab:SSD_architecture}. The initial goal of this evaluation was to evaluate all possible integrations of ZNS devices. For this we establish the following configuration: 

    \noindent \textbf{f2fs.} The f2fs (Figure~\ref{fig:ZNS_configurations}(c)) parameters are mostly kept at its default options, with the only change being the enforcing of 10\% overprovisioning. Since f2fs requires a randomly writable device for its metadata, we expose 4 of the zones on the ZNS device as randomly writable space, corresponding to the maximum amount that the ZNS device can expose as randomly writable space. This space is then used by f2fs for metadata and write caching. However, this requires the zoned space of f2fs (where the actual file system data will be) to align with the size of the randomly writable space, i.e., the randomly writable space has to be large enough to fit all the metadata for the file system. Therefore, the resulting largest possible size that successfully formats the f2fs file system on the zoned space is 100GiB. As a result, we create a 100GiB namespace from the available zoned capacity for the f2fs file system, which we utilize for all experiments. We do not use an additional larger randomly writable device, as we aim to avoid performance implications of multiple devices, which would make it difficult to differentiate between performance effects from the ZNS device and the additional block device.

\noindent \textbf{ZenFS.} ZenFS (Figure~\ref{fig:ZNS_configurations}(d)) requires an auxiliary path for its metadata to store LOG and LOCK files for RocksDB at runtime. With this it additionally allows to backup and recover a file system through its command line options, however we do not utilize this option in our evaluation. For the auxiliary path we use the 4GiB randomly writable space exposed by the ZNS device and create a f2fs file system on it which is mounted for solely the ZenFS auxiliary path to be placed on. 

As the largest zoned space that successfully formats f2fs is 100GiB, we utilize a 100GiB namespace for all experiments, alongside the 4GiB randomly writable namespace, and all remaining capacity is left in an unused namespace. Throughout the evaluation we refer to the 4GiB randomly writable space exposed by the ZNS device as the \textit{conventional namespace}, and the namespace containing the zoned storage on the ZNS device is referred to as the \textit{zoned namespace}. Lastly, since the ZNS specification was integrated in Linux Kernel 5.9.0+, we use a later Kernel version 5.12.0.

\section{Unsuccessful Experiments}\label{sec:unsuccessful_exps}
Designing of experiments to evaluate the performance of the varying levels of integration proved challenging, as initial experiments did not provide insightful results. We provide the iterations of experimental design and why experiments failed in an effort for others evaluating ZNS performance to avoid these pitfalls. Section~\ref{sec:failed_setup} provides the configurations of the various benchmark, followed by Section~\ref{sec:pitfalls_to_avoid} describing the failures of these benchmarks and pointing out possible causes for this. Lastly, we give additional lessons learned during this evaluation in Section~\ref{sec:lessons_learned}.

\subsection{Benchmark Setup}\label{sec:failed_setup}
We run several benchmarks to evaluate different performance aspects of ZNS devices. While each of the evaluations relies on db\_bench and RocksDB, they have slightly different benchmarking configurations. Below we describe the workload parameters for the different benchmarks.

\subsubsection{Integration Level}
We first benchmark the different possible integration levels, as shown in Figure~\ref{fig:ZNS_configurations}. Benchmarks initially fill the database in random and sequential key order, followed by reading from the database in random and sequential key order. We additionally run an overwrite benchmark, which overwrites existing keys in random order, and an updaterandom workload that modifies values of random keys. The overwrite benchmark and the updaterandom differ in the way values are accessed and modified. Overwrite issues asynchronous writes to the key, whereas updaterandom uses a read-modify-write approach. Keys are 16B each, values are configured to be 100B, and we disable any data compression. As mentioned, ZNS devices require direct I/O, we therefore set appropriate db\_bench flags to issue direct reads and use direct I/O for flushing and compaction.

\subsubsection{ZenFS Benchmark}\label{sec:zenfs_exp}
Next, to verify correctness of our setup we attempt to repeat an experiment depicted in~\cite{2021-Bjorling-ZNS}, comparing db\_bench performance on a configuration with f2fs to a configuration with ZenFS. The focus of this benchmark is write intensive workloads, as this is meant to trigger increased garbage collection and showcase gains of managing ZNS at the application-level with ZenFS. The benchmark is configured to run fillrandom and overwrite with a key size of 100B and value size of 800B. We additionally use data compression to compress data down to 400B. The original paper uses the entire device for its benchmark, however we scale it to the maximum possible that successfully fits into the namespace, which is equivalent to 50 million keys. Lastly, we set the target file size of SST files to be equal to the zone capacity, and again utilize appropriate flags for direct I/O.

\subsubsection{Multi-Tenancy}
Lastly, we run an experiment to evaluate how multiple concurrently running namespaces affect the performance of one another. For this we mount f2fs on the 100GiB zoned namespace and create an additional 100GiB namespace on which ZenFS with a db\_bench benchmark is running. We compare the performance of running the ZenFS namespace alone, without any interference from another namespace, to running the f2fs namespace concurrently. The ZenFS namespace runs the same workload as describe in the previous benchmark (Section~\ref{sec:zenfs_exp}). The goal being that the f2fs namespace creates substantial device traffic through write I/Os, especially during garbage collection from the overwrite benchmark, and thus show the performance impact on the ZenFS namespace.

\subsection{Pitfalls to Avoid}\label{sec:pitfalls_to_avoid}
Results of all experiments showed little to no performance difference in their benchmarks. Especially the ZenFS benchmark (Section~\ref{sec:zenfs_exp}), where the original paper showed substantial performance gains with ZenFS, in particular on overwrite benchmarks where GC is being triggered heavily. We failed to reproduce these exact results and only had minor performance gains from ZenFS. The multi-tenant evaluation showed very similar results, which appear contrary to prior expectations. That is if one namespace fully utilizes the device, then another namespace attempting to use the same device will have some performance implications, as they would now be sharing the device resources.

As the prior experiments proved ineffective in their evaluation, we propose the assumption that for performance
differences to appear on the ZNS device, it has to be largely utilized, such that LBA mappings are fully setup and more
garbage collection is triggered. If the device is not largely utilized, LBA mappings are not fully setup and the device
is able to provide peak performance without showing effects of garbage collection, as there is a large amount of free
space it can utilize. Additionally, the device bandwidth has to be utilized to the extent that it competes with the
garbage collection happening in the background. We therefore believe that with our setup we failed to produce enough
device utilization in bandwidth and space, and thus evaluations showed that there are no performance differences. While
we did not evaluate all configurations under increased garbage collection and device utilization, we believe it to have
an effect on performance and thus suggest its evaluation as future work (discussed in detail in
Section~\ref{sec:future_work}). Per suggestions of Western Digital, for significant performance advantages to appear for
ZNS devices, utilization of available storage capacity should be $>50\%$ (preferably $80\%+$) and bandwidth utilization
of $>30\%$ of the device bandwidth, such that the write workload competes with ongoing garbage collection. However, we
have not evaluated these specific configurations and thus leave it as future work.

\subsection{Lessons Learned}\label{sec:lessons_learned}
In addition to unsuccessful experiments we encountered several obstacles that were not immediately obvious to debug. Again, we provide our experiences in order for others to avoid these pitfalls. Firstly, by default the ZNS devices do not set a scheduler, neither do the applications such as f2fs or ZenFS, nor is there an error on an invalid scheduler being set. Thus setting up of the applications and formatting the ZNS device completes successfully, while as soon as writes are issued to the device I/O errors appear. Therefore, it is important to ensure the correct scheduler is always set, that is the \textit{mq-deadline} scheduler, and every namespace requires it to be set individually after creation. The majority of applications utilize multiple outstanding I/Os per zone, thus the \textit{mq-deadline} scheduler is required, as it will hold back I/Os such that only one outstanding I/O is submitted to the device at any point in time. If the application issues a single I/O synchronously, the scheduler could be left at the default configuration set to \textit{none} (default setting in Linux 5.12.0), however here again the application must enforce the sequential write constraint.

Secondly, device mapper support is not there yet. A device mapper implements a host-side FTL that makes the sequential write required zones randomly writable by exposing the zoned device as a conventional block device. It controls data placement, maintains data mappings, and runs garbage collection, just as the FTL on traditional flash storage. An existing implementation, such as \textit{dm-zoned}~\cite{2016-dmzoned-git} is ZBC and ZAC compliant but not compliant to the new concepts of zone capacity from the ZNS specification. The \textit{dm-zap}~\cite{2022-dmzap-git} device mapper aims to be ZNS compliant, however it is still a prototype and not fully functional with ZNS devices yet. As a result, we are currently not able to evaluate performance of one level of integration (Figure~\ref{fig:ZNS_configurations}(b)), requiring a future evaluation when the device mapper support is functional.

\section{Adapted Experimental Setup}\label{sec:adapted_setup}
\begin{figure*}[!ht]
    \centering
    \subfloat[]{\includegraphics[width=0.45\textwidth]{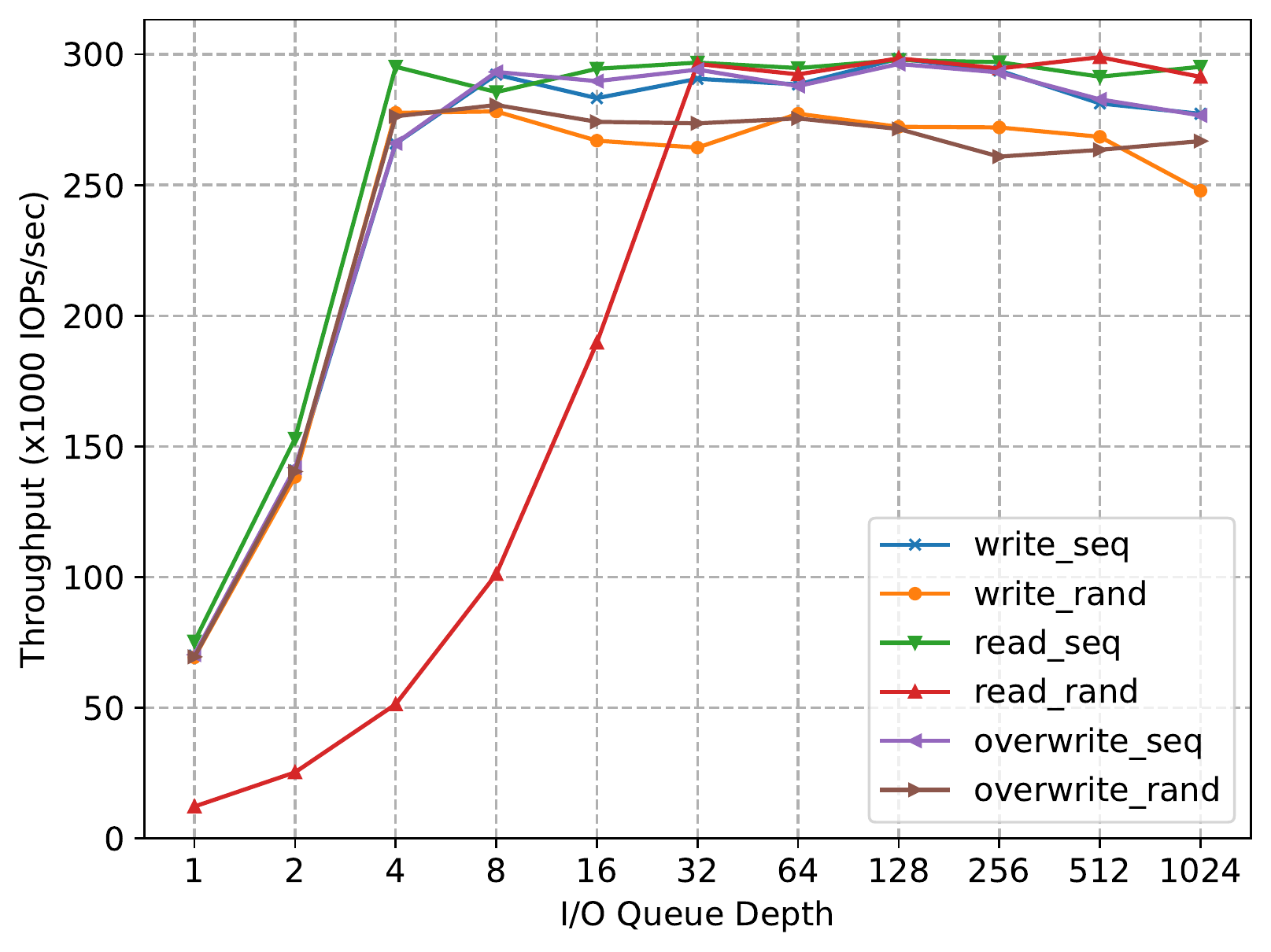}\label{fig:samsung_iops}}\hfill
    \subfloat[]{\includegraphics[width=0.45\textwidth]{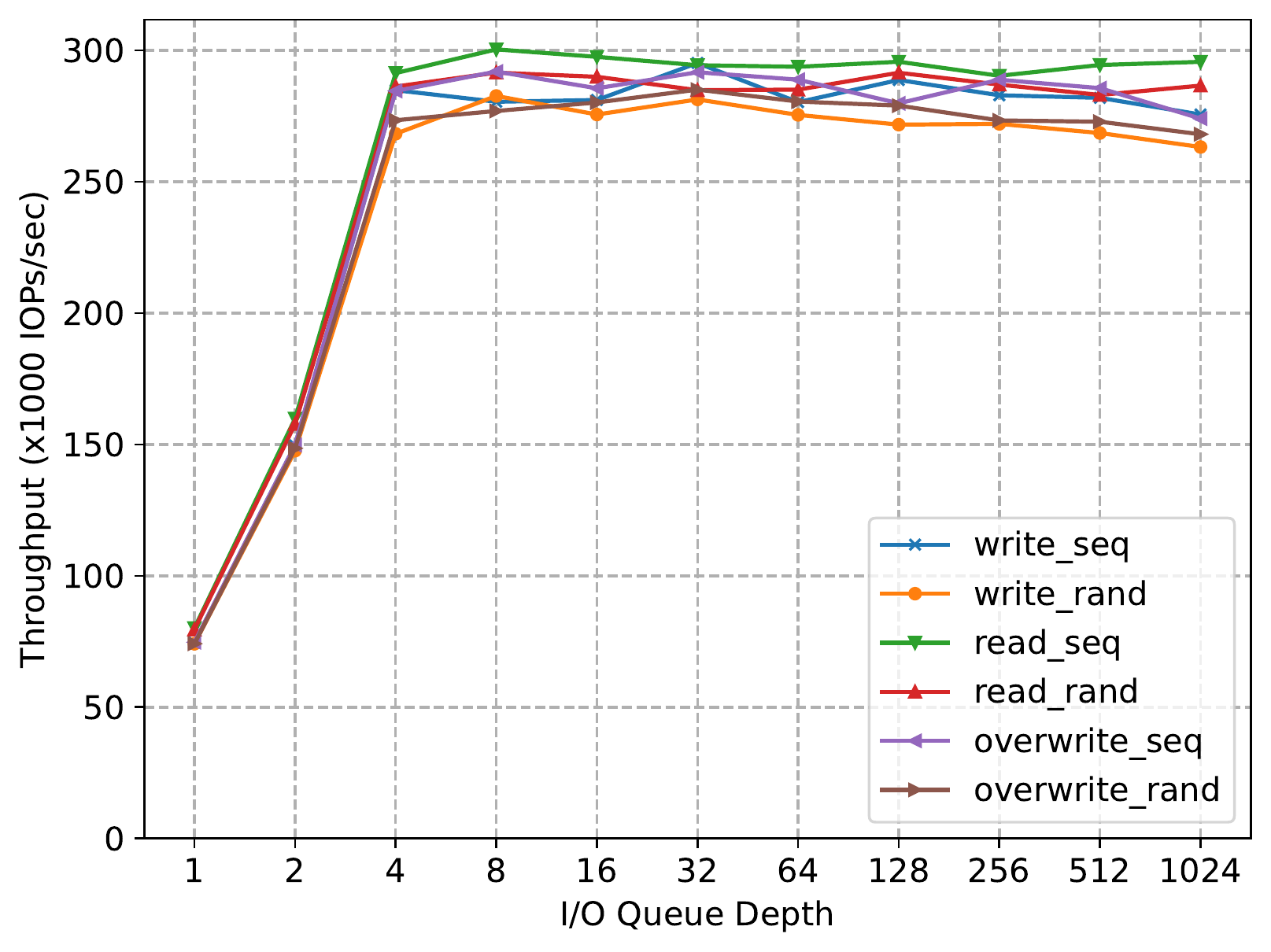}\label{fig:optane_iops}}
    \vspace{-0.3cm}
    \caption{Peak throughput of (a) the Samsung SSD and (b) the Optane SSD under various fio read and write workloads, with an increasing I/O queue depth (1-1024) and 4KiB block size.}
    \label{fig:conventional_iops}
\end{figure*}

Based on the prior failed experiments, we adapt the experimental setup by committing the unused space with a cold file in an effort to fully utilize the ZNS device, instead of leaving the unused space empty. We additionally make use of two conventional SSDs, one of which is a Samsung flash-based SSD and another Optane-based SSD. The conventional SSDs are used to provide a performance comparison between the conventional namespace on the ZNS device and regular SSDs. Detailed characteristics of the additional SSDs are presented in Table~\ref{tab:SSD_architecture}. 

The conventional SSDs do not support multiple namespaces, therefore separate partitions are used, with the same 100GB of experimental space, and the remaining space serving as storage for cold data. While the ZNS device used during the experiments supports setting sector sizes of 512B and 4KiB, we set the device to 512B, since the used conventional SSDs only support 512B sectors, and we aim to keep device configurations as similar as possible. Prior to running experiments devices are pre-conditioned to steady state performance by writing the entire space numerous times.

\section{Block-Level Device Performance}\label{sec:evaluation}
Focusing on the block-level performance of ZNS devices, we design several benchmarks using the fio benchmarking tool~\cite{fio-src}. In particular, we evaluate the following aspects of ZNS performance:

\begin{itemize}
    \item \textbf{ZNS block I/O performance for the conventional namespace (\cref{sec:block-level_ZNS_IO_Conv_ns}).} We establish the baseline performance of block I/O on the ZNS device for its conventional namespace, which is exposed as a small randomly writable space. We measure the achievable throughput and compare it to performance of conventional SSDs. Main findings show that, for achieving peak write bandwidth of the device larger block sizes are required.  
    \item \textbf{ZNS block I/O performance for the zoned namespace (\cref{sec:block-level_ZNS_IO_Zoned_ns}).} We measure the performance of the ZNS device, benchmarking its zoned namespace. Specifically, we identify the performance of the \textit{mq-deadline} and \textit{none} scheduler under various read and write workloads. Results show that sequential write performance is higher with the \textit{mq-deadline} scheduler, and read performance achieves lower median and tail latency with the scheduler set to \textit{none}.
\end{itemize}

\subsection{Conventional Device Performance}\label{sec:block-level_ZNS_IO_Conv_ns}

To measure the block-level I/O performance of the conventional namespace we run several fio workloads over the namespace and compare its performance to that of the Optane and Samsung SSDs.

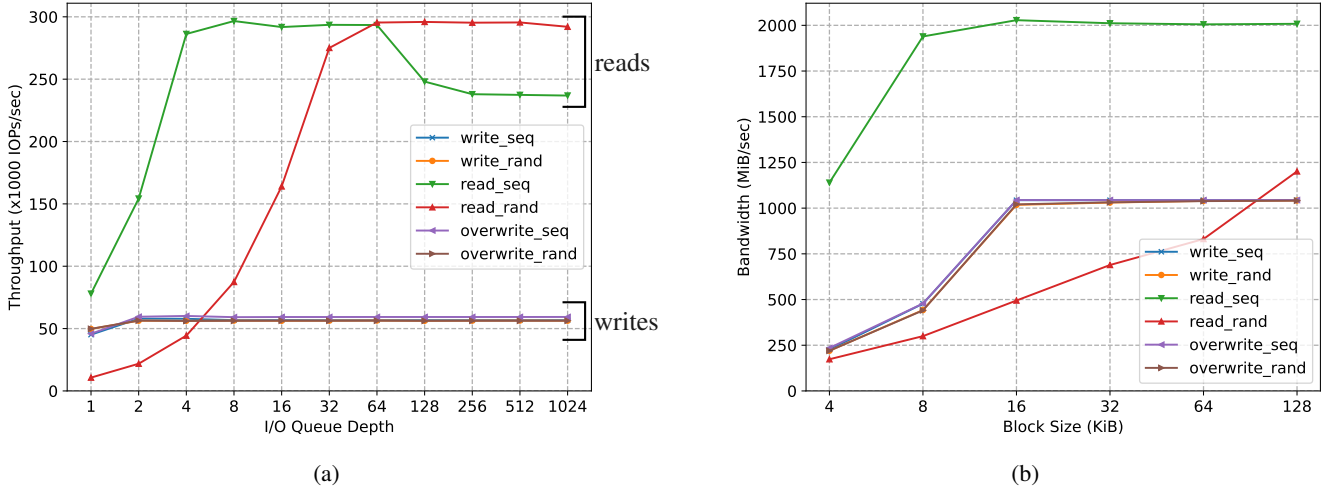
\begin{figure*}[!htb]
    \centering
    \subfloat[\label{fig:ZNS_conv_iops}]{
        \begin{tikzpicture}
            \node[anchor=south west, inner sep=0] at (2,0) {\includegraphics[width=0.45\textwidth]{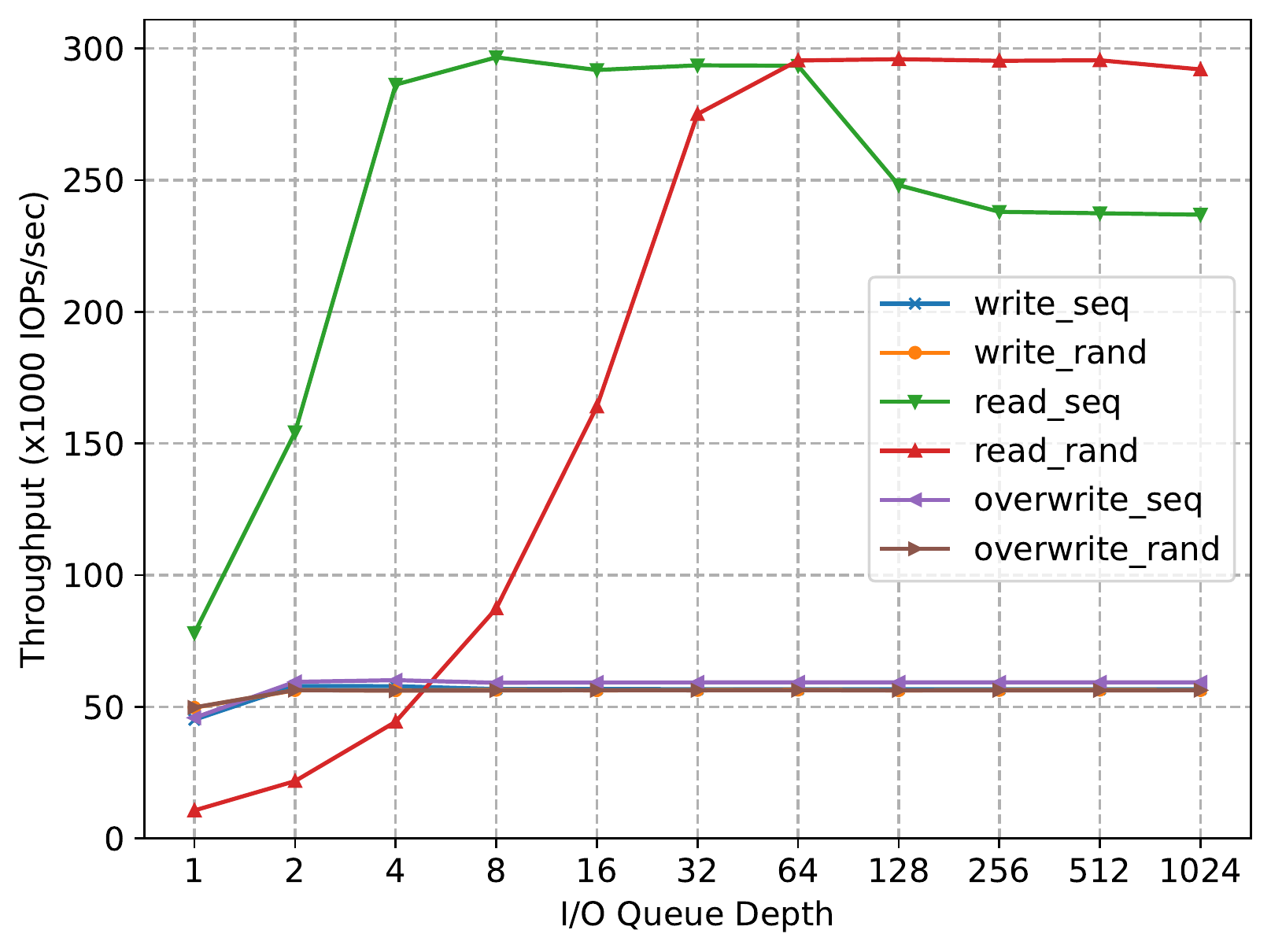}};
            \draw[black,thick] (9.5,5.7) -- (9.8,5.7) -- (9.8,4.5) -- (9.5,4.5);
            \node[font=\fontsize{8pt}{8pt}\color{black!90},anchor=center,text centered,align=center] at (10.3,5.1) {reads};
            \draw[black,thick] (9.5,1.9) -- (9.8,1.9) -- (9.8,1.4) -- (9.5,1.4);
            \node[font=\fontsize{8pt}{8pt}\color{black!90},anchor=center,text centered,align=center] at (10.35,1.65) {writes};
        \end{tikzpicture}}\hfill
    \subfloat[]{\includegraphics[width=0.45\textwidth]{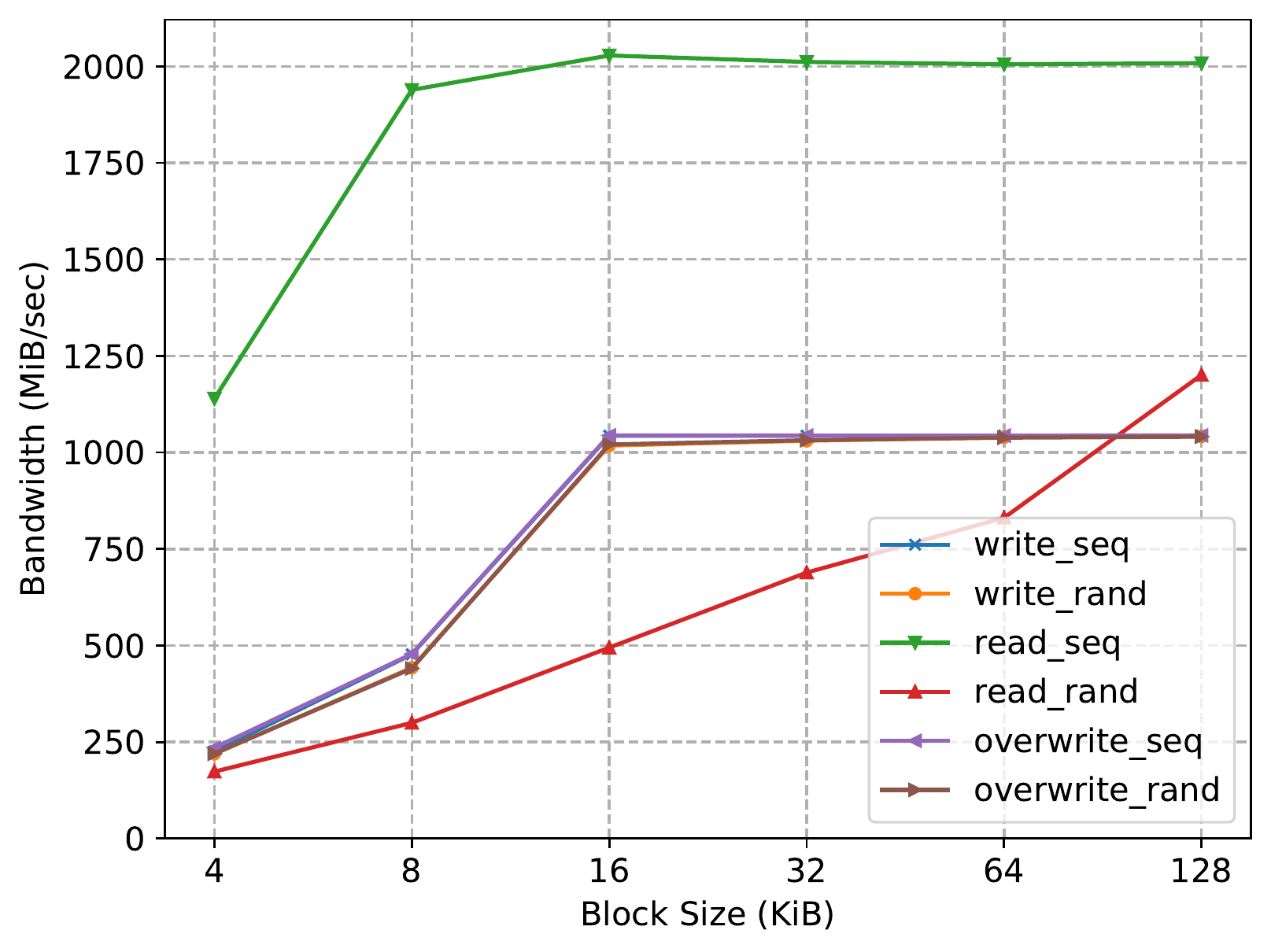}\label{fig:ZNS_scaled_bw}}
    \vspace{-0.3cm}
    \caption{Performance evaluation of the conventional namespace on the ZNS device with (a) the peak throughput under various fio read and write workloads, with an increasing I/O queue depth (1-1024) and 4KiB block size, and (b) the maximum achievable bandwidth with an I/O queue depth of 4 and an increasing block size (4KiB-128KiB).}
\end{figure*}

\noindent \textbf{Device throughput:} First, we measure the performance of the Optane and Samsung SSDs. For this, we run a fio benchmark that issues 4KiB read and write I/Os (I/O size is commonly referred to as the block size throughout this section). Specifically, we run the following benchmarks; sequential write, random write, sequential read, random read, sequential overwrite, and random overwrite. The overwrite benchmarks are achieved by fully writing the entire namespace and running a sequential and random write benchmark on the full namespace. Benchmarks are repeated with varying I/O queue depths, indicating the number of outstanding I/Os to maintain~\cite{2017-fio-documentation}. To avoid performance impact of NUMA effects on the results, we pin each workload to the NUMA node where the respective device is attached. 

Figures~\ref{fig:samsung_iops} and~\ref{fig:optane_iops} show the performance of the benchmarks for the Samsung SSD and the Optane based SSD, respectively. Both devices show a stable peak performance of 300KIOPs for small queue depths of 4, except for random reading on the Samsung SSD, which requires 16 outstanding I/Os to reach peak IOPs. Next, we run the same benchmarks on the conventional namespace exposed by the ZNS device. Figure~\ref{fig:ZNS_conv_iops} shows that the ZNS device only reaches a peak of 296KIOPs for read benchmarks at deeper queue depths of 8 and 64 for sequential and random reading, respectively. Write benchmarks reach peak performance at a shallow queue depth of 2, however performance is only 19\% of the peak device throughput of 296 KIOPs.

\noindent \textbf{ZNS device bandwidth:} As write performance for the conventional namespace is 81\% below the peak throughput of 296KIOPs, we additionally measure the achievable bandwidth for the ZNS device for larger block sizes. For this, we increase the block size to power of 2 values from 4KiB to 128KiB, and maintain a lower queue depth of 4. The resulting performance is depicted in Figure~\ref{fig:ZNS_scaled_bw}, showing an increase to a peak write bandwidth of 1GiB for block sizes from 16KiB and larger. Note, the throughput in Figure~\ref{fig:ZNS_conv_iops} reached peak performance of 296 KIOPS on sequential reading with a block size of 4KiB and a queue depth of 4, which is equivalent to $296\text{KIOPs}*4\text{KiB}=1.13\text{GiB}$. However, with an increasing block size the bandwidth increases to a peak of 2GiB for sequential reads. Unlike the sequential read performance, the write performance only reaches a peak bandwidth of 1GiB.

\noindent \textbf{Recommendations:} Based on this evaluation we can identify that for the particular ZNS device evaluated \textbf{(i)} sequential reading achieves a 94.2\% larger peak bandwidth than write performance at block size $\geq$ 16KiB, (\textbf{ii}) peak throughput of 296KIOPs for sequential reading is reached at lower queue depth of 4, while random reading requires deeper queues $\geq 64$ to achieve the same throughput, and \textbf{(iii)} for achieving peak write bandwidth of the device larger block sizes ($\geq$ 16KiB) are required.

\subsection{Zoned Device Performance}\label{sec:block-level_ZNS_IO_Zoned_ns}
These benchmarks focus purely on the zoned namespace performance, and quantify the overheads of using \textit{mq-deadline} and \textit{none} schedulers with ZNS devices. The \textit{mq-deadline} scheduler can utilize a higher I/O queue depth ($>1$), as it holds back I/Os and only submits a single I/O at a time, and with \textit{none} the host needs to ensure the I/O queue depth is equal to one.

\begin{figure*}[!t]
    \centering
    \subfloat[]{\includegraphics[width=0.45\textwidth]{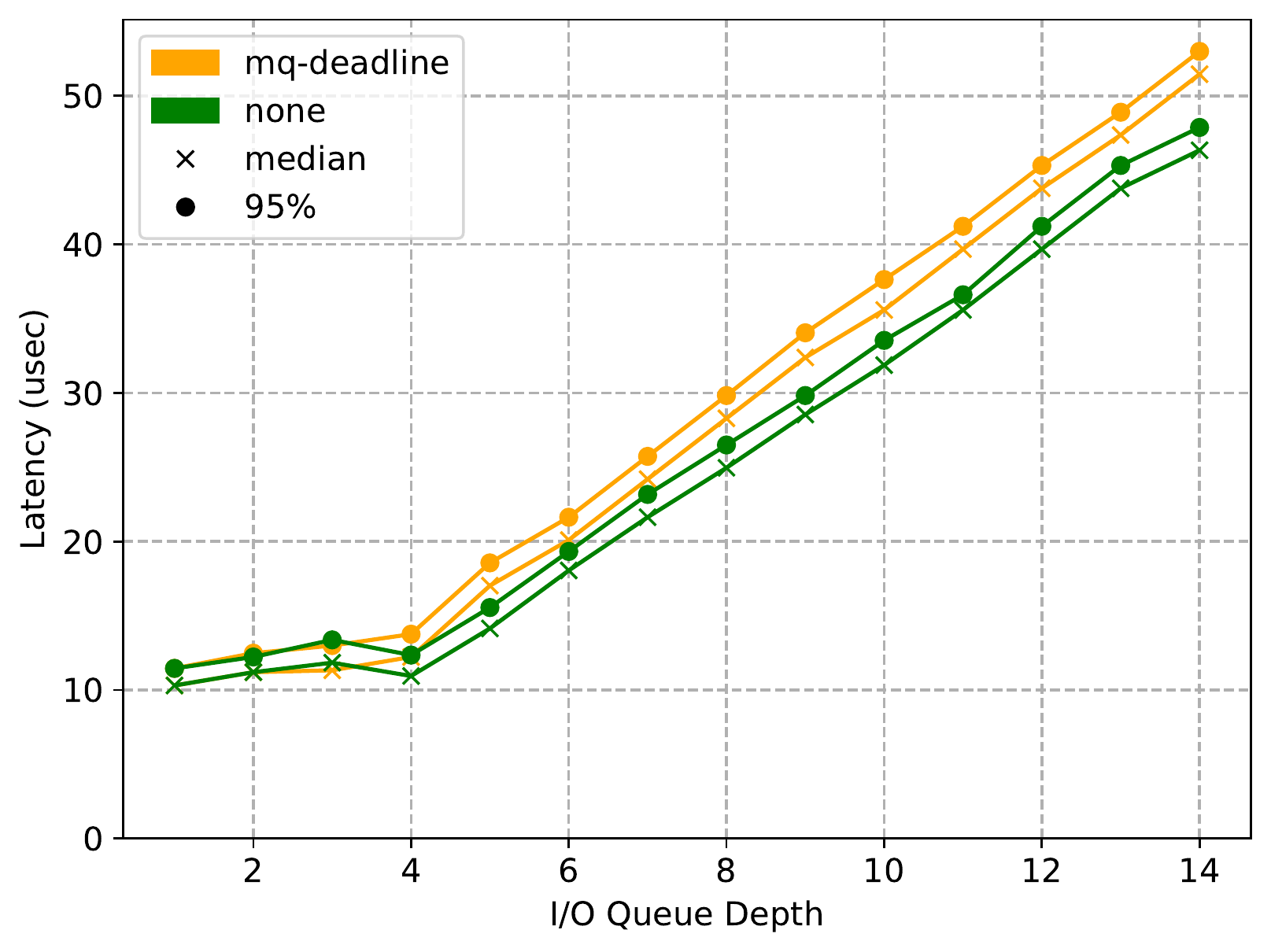}\label{fig:ZNS_concur_read_seq_iodepth}}\hfill
    \subfloat[]{\includegraphics[width=0.45\textwidth]{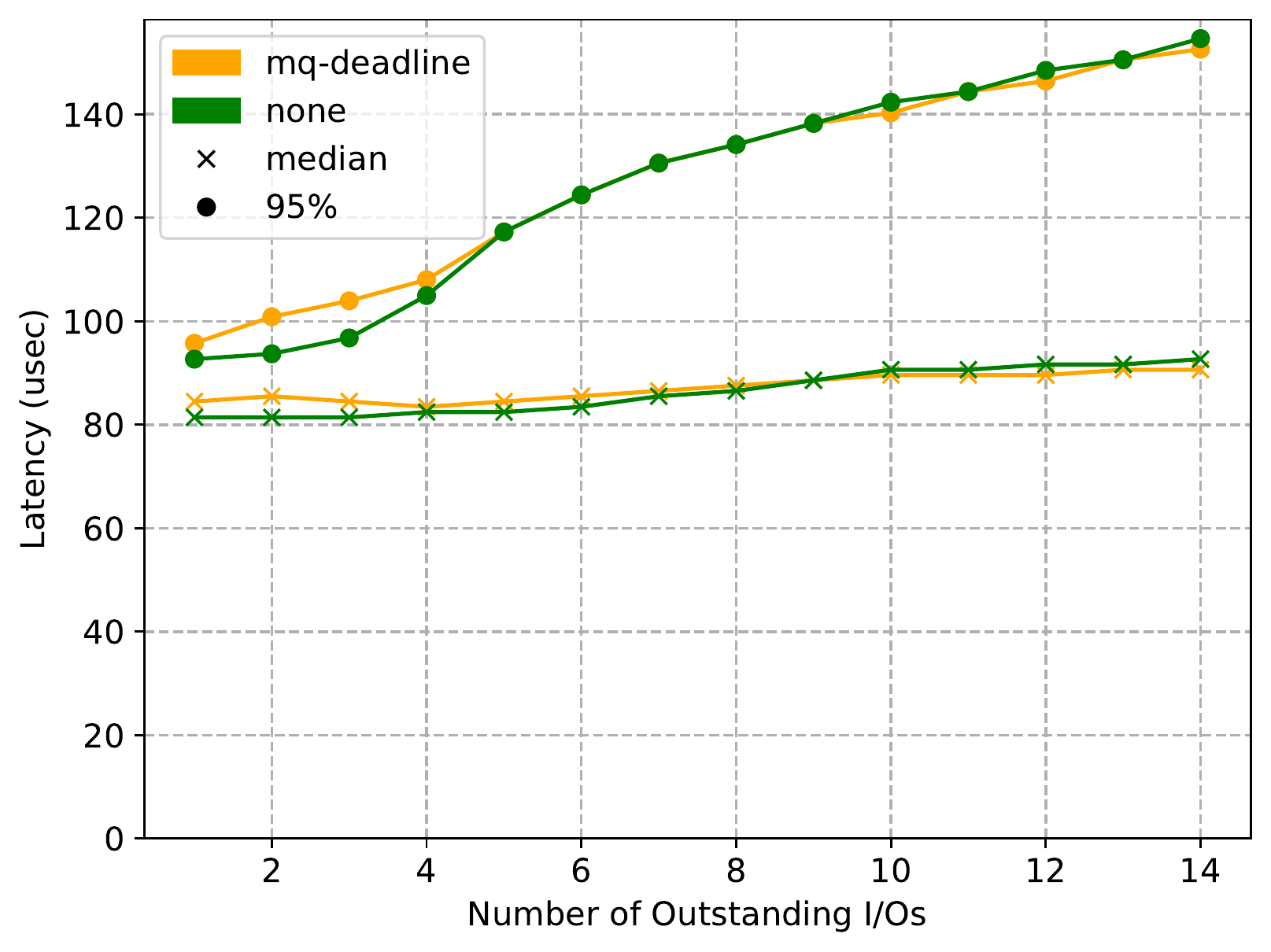}\label{fig:ZNS_concur_read_rand}}
    \vspace{-0.3cm}
    \caption{Latency of (a) issuing 4KiB sequential read I/Os in a single zone and increasing I/O queue depth (1-14) with \textit{mq-deadline} and \textit{none} schedulers, and (b) issuing 4KiB random read I/Os in a single zone with \textit{mq-deadline} and a single 4KiB random read I/O in each zone with increasing concurrent threads for the \textit{none} scheduler, giving both schedulers a certain number of outstanding random read I/Os (x-axis).}
\end{figure*}

\noindent \textbf{Read performance:} First, we measure the performance of sequential and random reading with both schedulers. Recall that ZNS does not enforce reading constraints, and thus both schedulers can have any number of outstanding read requests, with sequential or random accesses. The sequential read benchmark is configured to issue 4KiB read I/Os in a single zone, under both schedulers, and an increasing I/O queue depth, ranging from 1-14. Results presented in Figure~\ref{fig:ZNS_concur_read_seq_iodepth} show that the scheduler set to \textit{none} achieves a 9.95\% lower median and 9.66\% lower tail latency at an I/O queue depth of 14. This is due to the added overhead of having the \textit{mq-deadline} scheduler compared to \textit{none} bypassing the Linux I/O scheduler. 

Next, we benchmark random read performance. However, as sequential reading in a single zoned showed that bypassing the Linux I/O scheduler provides lower median and tail latency, we measure random reading by utilizing multiple zones for the \textit{none} scheduler. Specifically, \textit{mq-deadline} is set up with the same configuration as with sequential reading, issuing 4KiB I/Os in a single zone with an increasing I/O queue depth, while \textit{none} is set to issue a single I/O to a zone with an increasing number of concurrent threads (also ranging from 1-14, up to the maximum number of active zones). Thus, both schedulers have a particular number of outstanding I/Os (x-axis). Results shown in Figure~\ref{fig:ZNS_concur_read_rand} show that up to 4 outstanding I/Os \textit{mq-deadline} has a higher median latency, ranging between 1.23-4.79\% higher than median latency of \textit{none}. However, for more outstanding I/Os ($> 4$) both schedulers similar performance.

\begin{figure*}[!htb]
    \centering
    \subfloat[]{\includegraphics[width=0.48\textwidth]{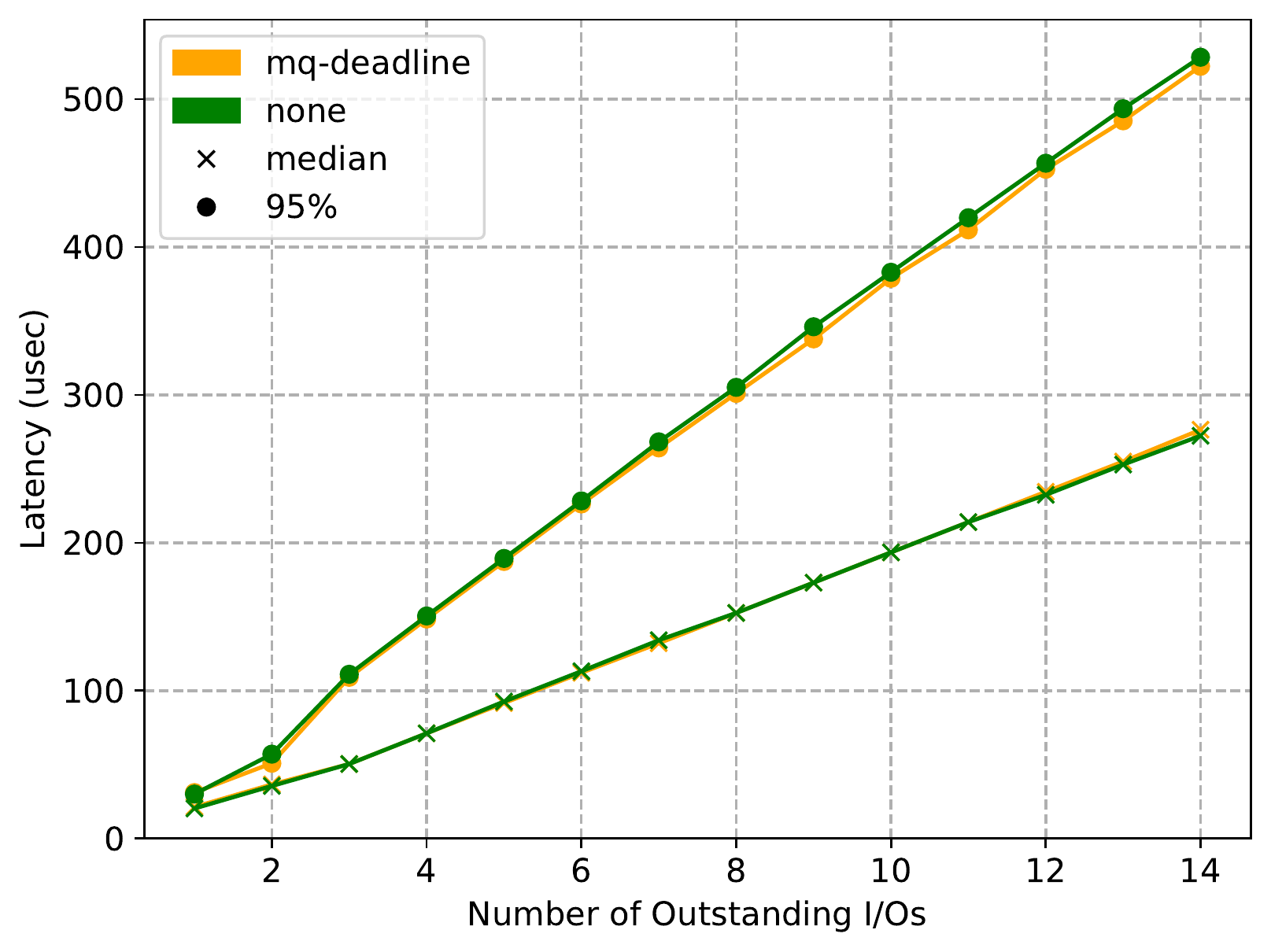}\label{fig:ZNS_concur_write_seq}}\hfill
    \subfloat[]{\includegraphics[width=0.48\textwidth]{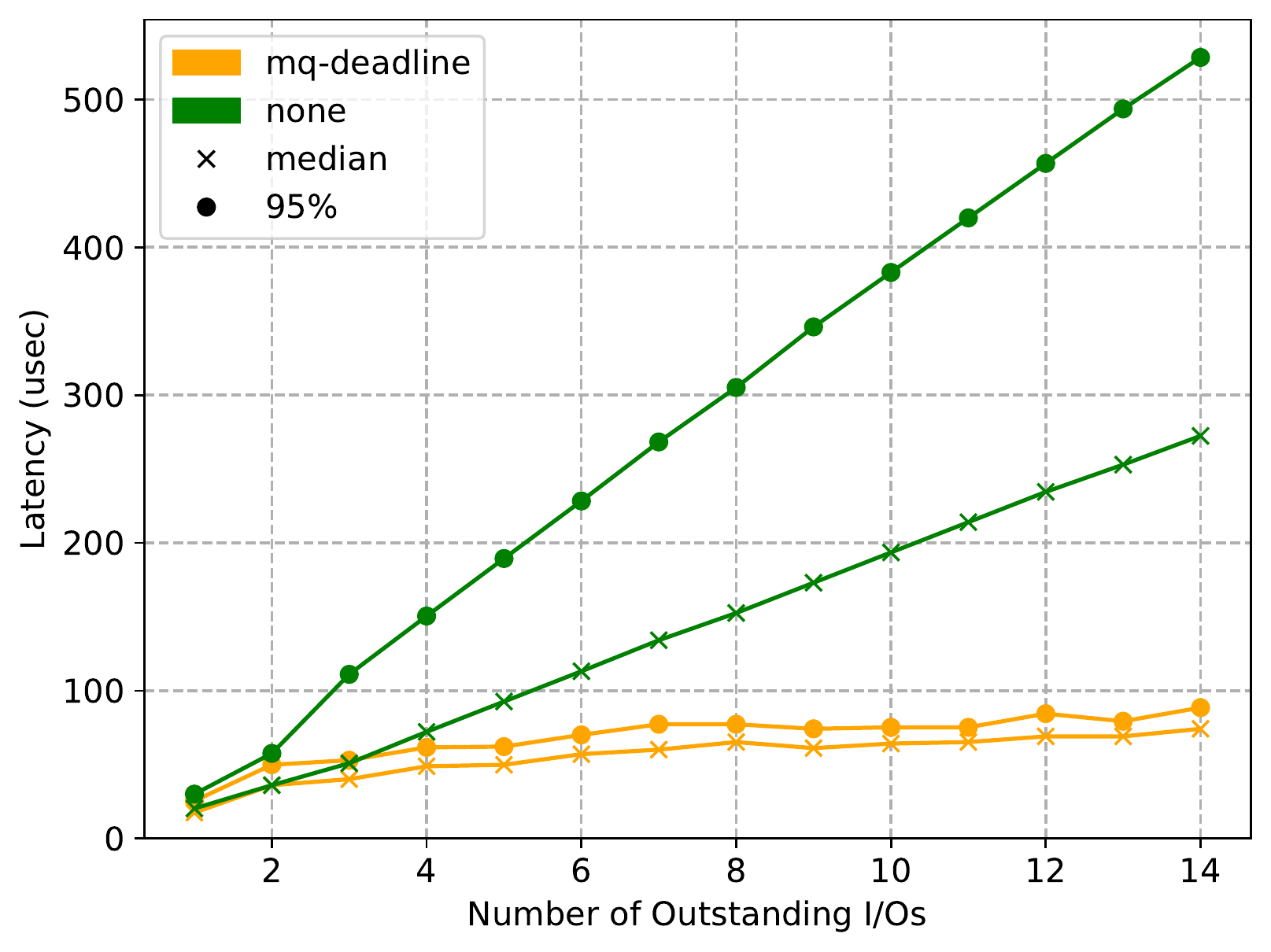}\label{fig:ZNS_concur_write_seq_iodepth}}
    \vspace{-0.3cm}
    \caption{Latency of (a) concurrently issuing 4KiB sequential write I/Os over an increasing number of active zones (and threads) and an I/O queue depth of 1 under \textit{mq-deadline} and \textit{none} scheduler, and (b) issuing 4KiB sequential write I/Os with increasing I/O queue depth (1-14) for \textit{mq-deadline} in a single zone, and I/O queue depth of 1 and increasing active zones (1-14) and concurrent threads with \textit{none} scheduler, giving both schedulers a certain number of outstanding write I/O requests (x-axis).}
\end{figure*}

\noindent \textbf{Write performance:} Next, we measure the performance of the ZNS device under write workloads with both schedulers. ZNS devices enforce write constraints, therefore only \textit{mq-deadline} can utilize I/O queue depths $> 1$. We first benchmark the performance of both schedulers by issuing a single 4KiB sequential write I/O in a zone with an increasing number of concurrent threads split across available zones on the device. The number of concurrent threads ranges from 1-14, equivalent to the maximum number of active zones on the device. Figure~\ref{fig:ZNS_concur_write_seq} shows that the resulting performance of both schedulers is nearly identical over all numbers of outstanding I/O requests.

Next, we measure the performance of \textit{mq-deadline} issuing I/Os in a single zone and increasing the I/O queue depth, rather than splitting individual I/Os concurrently across zones. The benchmark with the \textit{none} scheduler is the same as for the prior write benchmark, namely issuing 4KiB I/Os concurrently with an increasing number of threads split across the active zones. This allows both benchmarks to have a specific number of outstanding I/O requests (x-axis). Figure~\ref{fig:ZNS_concur_write_seq_iodepth} shows that the scheduler set to \textit{mq-deadline} achieves a 72.74\% lower median and 83.27\% lower tail latency. The lower latency for \textit{mq-deadline} as the number of outstanding I/Os increases is due to the merging of I/O requests into larger I/Os, as all the I/Os are at consecutive LBAs. This allows it to issue overall less I/Os than with \textit{none} scheduling in this configuration.

\noindent \textbf{Recommendations:} With this evaluation we can identify that \textbf{(i)} random read heavy workloads should avoid the Linux I/O scheduler by setting it to \textit{none}, providing up to 9.95\% lower median latency, and \textbf{(ii)} multiple outstanding write I/Os should utilize the \textit{mq-deadline} scheduler to merge I/Os in a single zone, rather than splitting I/Os concurrently over multiple zones.

\section{Related Work}\label{sec:related}
While ZNS has just recently been standardized~\cite{2021-Bjorling-ZNS}, there have been several initial evaluations and discussions of ZNS devices. Bjorling et al.~\cite{2021-Bjorling-ZNS} present modifications made to RocksDB and f2fs to support ZNS devices, and showcase the performance gains for these devices. Stavrinos et al.~\cite{2021-Stavrinos-Blockhead} provide an initial discussion on the benefits of ZNS devices and possible improvements for applications on them. Shin et al.~\cite{shin2020exploring} show a performance study of ZNS devices, depicting the need for large request sizes for achieving increased on-device parallelism. Similar to the results in our study, where the conventional namespace required larger block sizes ($\geq$ 16KiB) to reach peak device bandwidth. Han et al.~\cite{han2021zns+} provide a discussion on the ZNS interface, and propose an improved interface particularly optimized for log-structured file systems with segment compaction. However, there currently is no work that systematically studies the possible ways to integrate ZNS devices into the host software stack. We are the first to present a start at such a study.

As the ZBD model was originally introduced for SMR devices, there have been several performance evaluations of SMR devices and their integration into systems. Wu et al.~\cite{wu2017performance} showcase an extensive evaluation of host-aware SMR drives, which similarly to ZNS device expose device characteristics to the host. In particular, the authors focus on evaluating performance characteristics of the zoned interface. Additionally, Wu et al.~\cite{wu2016evaluating} provide a performance study on implications of different properties of host-aware SMR devices, including performance implications under the number of open zones. 

While not focusing on ZNS devices, past work presented similar evaluations for conventional SSDs, where He et al.~\cite{2017-He-SSD-Unwritten-Contract} provide an unwritten contract for flash-based SSDs, depicting numerous guidelines on performance improvements for such devices. Similarly, Wu et al.~\cite{2019-Wu-Unwritten_Contract_Optane} present such an unwritten contract for Optane SSDs. Both of these unwritten contracts were inspiration for us to provide a set of developer guidelines for the new ZNS devices. Yang et al.~\cite{2014-yang-dont_stack_log_on_log} characterize performance implications of building log-structured applications for flash-based SSDs, showcasing the negated benefits of optimizing application data structures for flash, caused by the device characteristics. Such an evaluation showcases application-level integration, which presents insightful results that should be reproduced on ZNS devices to further expand the developer guidelines we provide.

\section{Future Work}\label{sec:future_work}
With ZNS devices having just been introduced, they leave a plethora of avenues to explore. In particular, as we showcased in Figure~\ref{fig:ZNS_configurations}, there are numerous possibilities of integrating ZNS devices into the host software stack. In this evaluation we focus mainly on the block-level ZNS performance and provide several initial guidelines for developers. Expansion of these is left as future work by exploring the additional levels of integration. Specifically this includes evaluating the following aspects of ZNS devices:

\begin{itemize}
    \item What is the performance of the varying levels of integration for ZNS devices in terms of achievable IOPs, latency, transactions/sec, and additional instructions required? Under sequential and random, read and write workloads, are there performance implications of a specific integration, and what rules can be established when building applications for a particular integration?
    \item How is garbage collection influenced by the different levels of integration? Does building application specific garbage collection policies provide superior performance over other levels of integration? In particular how does an almost fully utilized device (e.g., 95\% utilization) affect garbage collection performance at the different levels?
    \item Do multiple applications running on the same ZNS device interfere with each other? This includes a shared device with multiple namespaces, shared block-level interface with for example multiple concurrently running file systems, and lastly a shared file system on ZNS with multiple concurrent applications. Does in each case an application's garbage collection impact the performance of concurrently running applications?
\end{itemize} 

As we evaluated some of these aspects and failed to produce insightful results, we propose to evaluate them by taking into account our shortcomings and considering the assumptions we provide in Section~\ref{sec:pitfalls_to_avoid}. An additional exploration that became apparent during this study was that there is currently no enforcing on the number of active zones at a time. The device only allows a maximum number of zones to be active at any point in time, however there is no managing of how active zones are split across namespaces on the device, or across applications running on the same namespace. Especially, as the device supports a larger number of namespaces to be created than active zones that can be open, a zone manager is required to assign zones across namespaces and applications, and provide fair resource sharing across namespaces and applications, enforcing that the maximum number is not exceeded, even if there are more concurrent applications running. 

Lastly, we suggest the exploration of file system improvements for f2fs, and other ZNS specific file systems, with grouping of files by creation time, death time, or owner as was suggested in~\cite{2021-Stavrinos-Blockhead} and similarly evaluated for conventional SSDs in~\cite{2017-He-SSD-Unwritten-Contract}. This would provide the possibility for optimizing garbage collection and improving the device performance at the file system integration level.

\section{Conclusion}\label{sec:conclusion}
The newly standardized ZNS device present a unique new addition to the host storage stack. Pushing the garbage collection responsibility up in the stack to the host allows for more optimal data placement, providing predictable garbage collection overheads, compared to overheads from conventional flash storage. We provide one of the first systematic studies analyzing the performance implications of ZNS integration, and present an initial set of development guidelines for ZNS devices. 

While we mainly focus on the block-level performance of ZNS devices, we additionally provide our unsuccessful experiments and pitfalls to avoid, and furthermore propose numerous future work ideas to evaluate ZNS device integration further and extend the guidelines provided in this work. Main findings in this evaluation show that sequential reads on ZNS devices achieve almost double the peak bandwidth of writes, and larger I/Os ($\ge$ 16KiB) are required for fully saturating the device bandwidth. Additionally, the selection of scheduler for ZNS devices can provide workload dependent performance gains.

\section*{Acknowledgments}
This work is generously supported by Western Digital (WD) donations. Matias Bjørling and the ZNS team at Western 
Digital provided many helpful and explanatory comments during the course of this study. We also thank Hans Holmberg for
providing comments and feedback on the report.
Animesh Trivedi is supported by the NWO grant number OCENW.XS3.030, Project Zero: Imagining a Brave CPU-free World! 

\section*{Availability}

All the collected data during this evaluation, scripts for running benchmarks, as well as plotting results are made publicly available at \url{https://github.com/nicktehrany/ZNS-Study}. Instructions and commands for usage of ZNS devices and reproducing of this evaluation are additionally provided in the Appendix. 

\bibliographystyle{plain}
\bibliography{main}

\begin{thebibliography}{10}

\bibitem{2008-Agrawal-Design-Tradeoff-SSD}
Nitin Agrawal, Vijayan Prabhakaran, Ted Wobber, John~D. Davis, Mark Manasse,
  and Rina Panigrahy.
\newblock {Design Tradeoffs for SSD Performance}.
\newblock In {\em USENIX 2008 Annual Technical Conference}, ATC'08, page
  57–70, USA, 2008. USENIX Association.

\bibitem{fio-src}
Jens Axboe.
\newblock {fio Flexible I/O Tester}.
\newblock \url{https://github.com/axboe/fio}.
\newblock Accessed: 2022-01-21.

\bibitem{2016-Axboe-f2fs_ZBC_patch}
Jens Axboe.
\newblock {[PATCH v8 0/7] ZBC / Zoned block device support}.
\newblock Patch, October 2016.
\newblock Available at:
  \url{https://www.mail-archive.com/linux-block@vger.kernel.org/msg01462.html}.

\bibitem{2017-fio-documentation}
Jens Axboe.
\newblock {fio - Flexible I/O tester rev. 3.27}.
\newblock Documentation, 2017.
\newblock Available at:
  \url{https://fio.readthedocs.io/en/latest/fio_doc.html}.

\bibitem{bjorling2020zoneappend}
Matias Bj{\o}rling.
\newblock {Zone Append: A New Way of Writing to Zoned Storage}.
\newblock February 2020.

\bibitem{2021-Bjorling-ZNS}
Matias Bj{\o}rling, Abutalib Aghayev, Hans Holmberg, Aravind Ramesh, Damien~Le
  Moal, Gregory~R. Ganger, and George Amvrosiadis.
\newblock {ZNS: Avoiding the Block Interface Tax for Flash-based SSDs}.
\newblock In {\em 2021 USENIX Annual Technical Conference (USENIX ATC 21)},
  pages 689--703. USENIX Association, July 2021.

\bibitem{2017-fast-lightnvm}
Matias Bj{\o}rling, Javier Gonzalez, and Philippe Bonnet.
\newblock {LightNVM}: The linux {Open-Channel} {SSD} subsystem.
\newblock In {\em 15th USENIX Conference on File and Storage Technologies (FAST
  17)}, pages 359--374, Santa Clara, CA, February 2017. USENIX Association.

\bibitem{2012-asplos-moneta}
Adrian~M. Caulfield, Todor~I. Mollov, Louis~Alex Eisner, Arup De, Joel Coburn,
  and Steven Swanson.
\newblock {Providing Safe, User Space Access to Fast, Solid State Disks}.
\newblock In {\em Proceedings of the Seventeenth International Conference on
  Architectural Support for Programming Languages and Operating Systems},
  ASPLOS XVII, page 387–400, New York, NY, USA, 2012. Association for
  Computing Machinery.

\bibitem{2014-ZBC}
INCITS T10~Technical Committee.
\newblock {Information technology - Zoned Block Commands (ZBC)}.
\newblock Standard, American National Standards Institute, September 2014.
\newblock Available from: \url{https://www.t10.org/}.

\bibitem{2015-ZAC}
INCITS T13~Technical Committee.
\newblock {Information technology – Zoned Device ATA Command Set (ZAC)}.
\newblock Standard, American National Standards Institute, December 2015.
\newblock Available from: \url{https://www.t13.org/}.

\bibitem{2022-dmzap-git}
Western~Digital Corporation.
\newblock {dm-zap}.
\newblock \url{https://github.com/westerndigitalcorporation/dm-zap}.
\newblock Accessed: 2022-01-21.

\bibitem{2022-zenfs-git}
Western~Digital Corporation.
\newblock {ZenFS: RocksDB Storage Backend for ZNS SSDs and SMR HDDs}.
\newblock \url{https://github.com/westerndigitalcorporation/zenfs}.
\newblock Accessed: 2022-01-21.

\bibitem{2021-ZNS-ZBD_docs}
Western~Digital Corporation.
\newblock {Zoned Block Device User Interface}.
\newblock Documentation.
\newblock Available at: \url{https://zonedstorage.io/docs/linux/zbd-api}.
  Accessed: 2022-05-02.

\bibitem{2021-ZNS-documentation}
Western~Digital Corporation.
\newblock {Zoned Storage - Write Ordering Control}.
\newblock Documentation.
\newblock Available at: \url{https://zonedstorage.io/docs/linux/sched}.
  Accessed: 2022-05-02.

\bibitem{2016-dmzoned-git}
Western~Digital Corporation.
\newblock {dm-zoned Device Mapper Userspace Tool}.
\newblock \url{https://github.com/westerndigitalcorporation/dm-zoned-tools},
  2016.
\newblock Accessed: 2022-01-21.

\bibitem{2013-Dean-tail_at_scale}
Jeffrey Dean and Luiz~Andr\'{e} Barroso.
\newblock {The Tail at Scale}.
\newblock {\em Commun. ACM}, 56(2):74–80, February 2013.

\bibitem{2014-Desnoyers-Analytic_Models_SSD}
Peter Desnoyers.
\newblock {Analytic Models of SSD Write Performance}.
\newblock {\em ACM Trans. Storage}, 10(2), March 2014.

\bibitem{2017-Dong-Optimizing-RocksDB}
Siying Dong, Mark Callaghan, Leonidas Galanis, Dhruba Borthakur, Tony Savor,
  and Michael Strum.
\newblock {Optimizing Space Amplification in RocksDB.}
\newblock In {\em CIDR}, volume~3, page~3, 2017.

\bibitem{Feldman2013ShingledMR}
Timothy~R. Feldman and Garth~A. Gibson.
\newblock {Shingled Magnetic Recording: Areal Density Increase Requires New
  Data Management}.
\newblock {\em login Usenix Mag.}, 38, 2013.

\bibitem{Gibson2011PrinciplesOO}
Garth Gibson and Greg Ganger.
\newblock {Principles of Operation for Shingled Disk Devices}.
\newblock In {\em 3rd Workshop on Hot Topics in Storage and File Systems
  (HotStorage 11)}, Portland, OR, June 2011. USENIX Association.

\bibitem{2009-Gupta-DFTL}
Aayush Gupta, Youngjae Kim, and Bhuvan Urgaonkar.
\newblock {DFTL: A Flash Translation Layer Employing Demand-Based Selective
  Caching of Page-Level Address Mappings}.
\newblock {\em SIGARCH Comput. Archit. News}, 37(1):229–240, March 2009.

\bibitem{han2021zns+}
Kyuhwa Han, Hyunho Gwak, Dongkun Shin, and Jooyoung Hwang.
\newblock {ZNS+: Advanced Zoned Namespace Interface for Supporting In-Storage
  Zone Compaction}.
\newblock In {\em 15th {USENIX} Symposium on Operating Systems Design and
  Implementation ({OSDI} 21)}, pages 147--162. {USENIX} Association, July 2021.

\bibitem{2017-He-SSD-Unwritten-Contract}
Jun He, Sudarsun Kannan, Andrea~C. Arpaci-Dusseau, and Remzi~H. Arpaci-Dusseau.
\newblock {The Unwritten Contract of Solid State Drives}.
\newblock In {\em Proceedings of the Twelfth European Conference on Computer
  Systems}, EuroSys '17, page 127–144, New York, NY, USA, 2017. Association
  for Computing Machinery.

\bibitem{2022-intel-p}
Intel.
\newblock {Intel® SSD D7-P5600 Series}.
\newblock
  \url{https://ark.intel.com/content/www/us/en/ark/products/202708/intel-ssd-d7p5600-series-6-4tb-2-5in-pcie-4-0-x4-3d3-tlc.html},
  Accessed: 2022-05-02.

\bibitem{2009-fast-dfs}
William~K. Josephson, Lars~A. Bongo, Kai Li, and David Flynn.
\newblock {DFS: A File System for Virtualized Flash Storage}.
\newblock {\em ACM Trans. Storage}, 6(3), sep 2010.

\bibitem{2014-hotstorage-multistream-ssd}
Jeong-Uk Kang, Jeeseok Hyun, Hyunjoo Maeng, and Sangyeun Cho.
\newblock {The Multi-Streamed Solid-State Drive}.
\newblock In {\em Proceedings of the 6th USENIX Conference on Hot Topics in
  Storage and File Systems}, HotStorage'14, page~13, USA, 2014. USENIX
  Association.

\bibitem{2015-Kim-SLO_complying_ssds}
Jaeho Kim, Donghee Lee, and Sam~H. Noh.
\newblock {Towards {SLO} Complying {SSDs} Through {OPS} Isolation}.
\newblock In {\em 13th USENIX Conference on File and Storage Technologies (FAST
  15)}, pages 183--189, Santa Clara, CA, February 2015. USENIX Association.

\bibitem{2015-Changman-f2fs}
Changman Lee, Dongho Sim, Jooyoung Hwang, and Sangyeun Cho.
\newblock {F2FS: A New File System for Flash Storage}.
\newblock In {\em 13th USENIX Conference on File and Storage Technologies (FAST
  15)}, pages 273--286, Santa Clara, CA, February 2015. USENIX Association.

\bibitem{2022-Rocksdb-src}
Facebook RocksDB.
\newblock {RocksDB: A Persistent Key-Value Store for Flash and RAM Storag},
  2022.
\newblock Available at: \url{https://rocksdb.org/}.

\bibitem{2022-samsung-zand}
Samsung.
\newblock {Ultra-Low Latency with Samsung Z-NAND SSD}.
\newblock
  \url{https://semiconductor.samsung.com/resources/brochure/Ultra-Low\%20Latency\%20with\%20Samsung\%20Z-NAND\%20SSD.pdf},
  Accessed: 2022-05-02.

\bibitem{shin2020exploring}
Hojin Shin, Myounghoon Oh, Gunhee Choi, and Jongmoo Choi.
\newblock {Exploring performance characteristics of ZNS SSDs: Observation and
  implication}.
\newblock In {\em 2020 9th Non-Volatile Memory Systems and Applications
  Symposium (NVMSA)}, pages 1--5. IEEE, 2020.

\bibitem{2021-Stavrinos-Blockhead}
Theano Stavrinos, Daniel~S. Berger, Ethan Katz-Bassett, and Wyatt Lloyd.
\newblock {Don't Be a Blockhead: Zoned Namespaces Make Work on Conventional
  SSDs Obsolete}.
\newblock In {\em Proceedings of the Workshop on Hot Topics in Operating
  Systems}, HotOS '21, page 144–151, New York, NY, USA, 2021. Association for
  Computing Machinery.

\bibitem{Suresh2012ShingledMR}
Anand Suresh, Garth~A. Gibson, and Gregory~R. Ganger.
\newblock {Shingled Magnetic Recording for Big Data Applications}.
\newblock 2012.

\bibitem{2022-nvme-spec}
NVM~Express Workgroup.
\newblock {NVM Express NVM Command Set Specification 2.0}.
\newblock Standard, January 2022.
\newblock Available from: \url{https://nvmexpress.org/specifications}.

\bibitem{wu2017performance}
Fenggang Wu, Ziqi Fan, Ming-Chang Yang, Baoquan Zhang, Xiongzi Ge, and David~HC
  Du.
\newblock {Performance evaluation of host aware shingled magnetic recording
  (HA-SMR) drives}.
\newblock {\em IEEE Transactions on Computers}, 66(11):1932--1945, 2017.

\bibitem{wu2016evaluating}
Fenggang Wu, Ming-Chang Yang, Ziqi Fan, Baoquan Zhang, Xiongzi Ge, and
  David~H.C. Du.
\newblock {Evaluating Host Aware {SMR} Drives}.
\newblock In {\em 8th USENIX Workshop on Hot Topics in Storage and File Systems
  (HotStorage 16)}, Denver, CO, June 2016. USENIX Association.

\bibitem{2019-Wu-Unwritten_Contract_Optane}
Kan Wu, Andrea Arpaci-Dusseau, and Remzi Arpaci-Dusseau.
\newblock {Towards an Unwritten Contract of Intel Optane {SSD}}.
\newblock In {\em 11th USENIX Workshop on Hot Topics in Storage and File
  Systems (HotStorage 19)}, Renton, WA, July 2019. USENIX Association.

\bibitem{2014-yang-dont_stack_log_on_log}
Jingpei Yang, Ned Plasson, Greg Gillis, Nisha Talagala, and Swaminathan
  Sundararaman.
\newblock {Don{\textquoteright}t Stack Your Log On My Log}.
\newblock In {\em 2nd Workshop on Interactions of NVM/Flash with Operating
  Systems and Workloads (INFLOW 14)}, Broomfield, CO, October 2014. USENIX
  Association.

\end{thebibliography}

\appendix
\section{General Information}
Throughout this guide, commands and set up explanation contain names of the specific NVMe device which are depicting their configuration in our system. \textit{nvme0n1p1} depicts the Samsung SSD, and \textit{nvme1n1p2} the Optane SSD, both of which only support a single namespace, requiring a partition to set up the 100GB experimental space. The conventional namespace of the ZNS device is \textit{nvme2n1} and the zoned namepace is \textit{nvme2n2}. We provide scripts for automation of all these benchmarks~\footnote{Available at \url{https://github.com/nicktehrany/ZNS-Study}}, as often numerous steps and retrieval of device specific information is required.

\section{Device Setup}\label{appendix:Device_Setup}
This section contains all required setup for devices, including namespace configuration, as well as the required applications for the different configurations used in this evaluation. To interact with ZNS devices, libnvme~\footnote{Available at \url{https://github.com/linux-nvme/libnvme}}, nvme-cli~\footnote{Available at \url{https://github.com/linux-nvme/nvme-cli}}, blkzone from util-linux~\footnote{Available at \url{https://github.com/util-linux/util-linux}}, libzbd~\footnote{Available at \url{https://github.com/westerndigitalcorporation/libzbd}}, and all their dependencies need to be installed. Note, ZNS integration is largely still new to applications used in this evaluation, therefore using the master branch is often required. Additionally, ZNS support was added to the Linux Kernel 5.9, therefore this version or newer one is required. 

\subsection{ZNS Device Configuration}
As we are using several namespaces on the ZNS device, one to expose a small amount of conventional randomly writable area, another of 100GiB (50 zones), and one for the remaining available space. The command for identifying the available size of the device is shown in listing~\ref{Listing:ZNS_info}, note that the ZNS device is configured to a 512B sector size. See Appendix~\ref{appendix:ZNS_info} on how to retrieve the supported sector sizes for a device. Next, the creation of namespaces is depicted in listing~\ref{Listing:ZNS_setup}. After creation of the namespaces, it is important to set the appropriate scheduler for all zones namespaces, as applications often do not do this automatically or check if the desired scheduler is set.

\begin{lstlisting}[language=bash,caption={Setting up NVMe ZNS namespaces.},label={Listing:ZNS_setup}]
# Create a 100GiB namespace, size is given in 512B sectors
$ sudo nvme create-ns /dev/nvme2 -s 209715200 -c 209715200 -b 512 --csi=2
# Repeat for all namespaces with according size
# Attach all namespaces to same controller (adapt -n argument with ns id)
sudo nvme attach-ns /dev/nvme2 -n 1 -c 0

# Set the correct scheduler for all zoned namespaces (adapt device path for each ns)
$ echo mq-deadline | sudo tee /sys/block/nvme2n2/queue/scheduler
\end{lstlisting}

\subsection{f2fs Configuration}
Setting up of f2fs requires f2fs-tools~\footnote{Available at \url{https://git.kernel.org/pub/scm/linux/kernel/git/jaegeuk/f2fs-tools.git/about/}} to make the file system. Configurations of f2fs with a ZNS device require an additional regular block device that is randomly writable, due to f2fs using in place updates for metadata and the superblock. In addition, both devices have to be configured to the same sector size. The exact commands for creating of the f2fs file system and mounting it are shown in listing~\ref{Listing:f2fs}. The order of devices specified lists one or more zoned device, followed by a single conventional block device that is randomly writable. The location of the superblock is used for mounting, hence only the randomly writable device used for file system creating is provided in the mount command. 

\begin{lstlisting}[language=bash,caption={Creating and mounting of f2fs file system. Requires the ZNS device and an additional randomly writable block device.},label={Listing:f2fs}]
# Format devices and create fs
$ sudo mkfs.f2fs -f -m -c /dev/nvme2n2 /dev/nvme2n1
$ sudo mount -t f2fs /dev/nvme2n1 /mnt/f2fs/
\end{lstlisting}

\subsection{ZenFS Configuration}
ZenFS~\footnote{Available at \url{https://github.com/westerndigitalcorporation/zenfs}} provides the storage backend for RocksDB to provide usage of ZNS devices. The ZenFS file system allows to be backed up and recovered to avoid data loss in failure events. Setting up of ZenFS requires the zoned device and an additional auxiliary path on another device with a file system, where it places the backup files, as well as any LOG and LOCK files required during RocksDB runtime. The command to set up ZenFS on a zoned device is shown in listing~\ref{Listing:zenfs}. The auxiliary path is placed on a conventional block device that is mounted with f2fs. 

\begin{lstlisting}[language=bash,caption={Creating of ZenFS file system. Requires an auxiliary path to place metadata.},label={Listing:zenfs}]
$ sudo ./plugin/zenfs/util/zenfs mkfs --zbd=nvme2n2 --aux_path=/home/nty/rocksdb_aux_path/zenfs2n2
\end{lstlisting}

\subsection{Namespace initialization}
As mentioned previously, we utilize a 100GiB namespace (\textit{nvme2n2}) for experiments and leave the remaining available space in a separate namespace (\textit{nvme2n3}) to be filled with cold data. Listing~\ref{Listing:ZNS_Cold_Data_NS} shows how this is achieved using fio~\footnote{Available at \url{https://github.com/axboe/fio}}. For fio to be able to write the entire namespace on the device, it requires the block size to be a multiple of the zone capacity (see listing~\ref{Listing:ZNS_info} on how to retrieve it). Similarly, the conventional devices (Optane and Samsung SSDs) also have their free space filled with cold data with the second shown command. The command is set up to write 2TiB however, fio will quit once the device is full. Additionally, note that as mentioned earlier the conventional SSDs only support a single namespace and hence have separate partitions set up for the experimental and cold data space.

\begin{lstlisting}[language=bash,caption={Filling namespace 3 with cold data.},label={Listing:ZNS_Cold_Data_NS}]
# Fill ZNS free space with cold data
$ sudo fio --name=zns-fio --filename=/dev/nvme2n3 --direct=1 --size=$((4194304*512*`cat /sys/block/nvme2n3/queue/nr_zones`)) --ioengine=libaio --zonemode=zbd --iodepth=8 --rw=write --bs=512K

# Fill conventional SSD free space with cold data
$ sudo fio --name=zns-fio --filename=/dev/nvme0np2 --direct=1 --size=2T --ioengine=libaio --iodepth=8 --rw=write --bs=512K
\end{lstlisting}

\section{Getting ZNS Device Information}\label{appendix:ZNS_info}
There are several attributes to the ZNS device that are required for later experiments, such as the zone capacity and the number of allowed active zones. Listing~\ref{Listing:ZNS_info} illustrates how to retrieve these. Supported sector sizes can be checked with the provided command, and are presented in powers of 2. Hence, a \textit{lbads:9} is equivalent to $2^9=512$ Bytes. We additionally retrieve the NUMA node at which the device is attached to, in order to pin workloads to this specific NUMA node.

\begin{lstlisting}[language=bash,caption={Retrieving information about the ZNS device.},label={Listing:ZNS_info}]
# Get the available device capacity in 512B sectors
$ sudo nvme id-ctrl /dev/nvme2 | grep tnvmcap | awk '{print $3/512}'

# Get the zone capacity in MiB 
$ sudo nvme zns report-zones /dev/nvme2n2 -d 1 | grep -o 'Cap:.*$' | awk '{print strtonum($2)*512/1024/1024}'

# Get maximum supported active zones
$ cat /sys/block/nvme2n2/queue/max_active_zones

# Get the supported sector sizes in powers of 2
$ sudo nvme id-ns /dev/nvme2n2 | grep -o "lbads:[0-9]*"

# Get NUMA node device is attached at
$ cat/sys/block/nvme2n1/device/numa_node
\end{lstlisting}

\section{ZNS Block-Level I/O Performance}
This section contains the commands used to establish the baseline maximum performance of the device, as well as extracted metrics for comparison to later experiments. All experiments for this are using fio as benchmarking tool. Appendix~\ref{appendix:Conv_device_performance} shows the set-up and commands for the provided evaluation in \cref{sec:block-level_ZNS_IO_Conv_ns}, and Appendix~\ref{appendix:Zoned_device_performance} shows the same evaluation on the ZNS device, as presented in \cref{sec:block-level_ZNS_IO_Zoned_ns}.

\subsection{Conventional Device Performance}\label{appendix:Conv_device_performance}
\noindent \textbf{Device throughput:} We run sequential and random writing and reading benchmarks for a block size (I/O size) of 4KiB under varying I/O queue depths to identify the maximum achievable IOPs of the devices. We run this on all three devices, since they all expose the regular conventional block device interface without write constraints, hence we only use the conventional namespace of the ZNS device in this section. The commands are shown in Listing~\ref{listing:conv_performance}.

The order of benchmarks is intentional such that randomwrite first runs, then the namespace is reset and written with the sequential benchmark. Overwrite benchmarks are done similarly with sequential and random writing after the entire namespace is filled with a write benchmark. Note that benchmarks are pinned to the NUMA node where the device is attached (see Appendix~\ref{appendix:ZNS_info} for retrieval of this). All defined variables (indicated by the \$ before a variable) are set up by our script, however for manual running of these commands require to simply be replaced by their associated value. The name and output argument depict naming for our plotting script to parse, however can simply be changed.

\noindent \textbf{ZNS device bandwidth:} We additionally showed the bandwidth scaling for the conventional namespace on the ZNS device for a I/O queue depth of 4 and increasing block sizes, which are shown in Listing~\ref{listing:conv_bandwidth}. Again, replace all defined variables with their respective value.

\begin{lstlisting}[language=bash,caption={Establishing the peak IOPs for the conventional devices with 4KiB block size and varying I/O queue depths.},label={listing:conv_performance}]
# We define several variables, replace these with values
    # DEV: device name (e.g., nvme2n1)
    # depth: I/O queue depth (from 1-1024 in powers of 2)
    # DEV_NUMA_NODE: NUMA Node of the device
    # SIZE: device size (e.g., 100G)

$ sudo numactl -m $DEV_NUMA_NODE fio --name=$DEV_randwrite_4Ki_queue-depth-$depth --output=$DEV_randwrite_4Ki_queue-depth-$depth.json --output-format=json --filename=/dev/$DEV --direct=1 --size=$SIZE --ioengine=libaio --iodepth=$depth --rw=randwrite --bs=4Ki --runtime=30 --numa_cpu_nodes=$DEV_NUMA_NODE --ramp_time=10 --time_based --percentile_list=50:95

# Reset the namespace between write benchmarks (only on namespaces, not partitions)
$ sudo nvme format /dev/$DEV -f

# Run remaining benchmarks
$ sudo numactl -m $DEV_NUMA_NODE fio --name=$DEV_write_4Ki_queue-depth-$depth --output=$DEV_write_4Ki_queue-depth-$depth.json --output-format=json --filename=/dev/$DEV --direct=1 --size=$SIZE --ioengine=libaio --iodepth=$depth --rw=write --bs=4Ki --runtime=30 --numa_cpu_nodes=$DEV_NUMA_NODE --ramp_time=10 --time_based --percentile_list=50:95
$ sudo numactl -m $DEV_NUMA_NODE fio --name=$DEV_read_4Ki_queue-depth-$depth --output=$DEV_read_4Ki_queue-depth-$depth.json --output-format=json --filename=/dev/$DEV --direct=1 --size=$SIZE --ioengine=libaio --iodepth=$depth --rw=read --bs=4Ki --runtime=30 --numa_cpu_nodes=$DEV_NUMA_NODE --ramp_time=10 --time_based --percentile_list=50:95
$ sudo numactl -m $DEV_NUMA_NODE fio --name=$DEV_randread_4Ki_queue-depth-$depth --output=$DEV_randread_4Ki_queue-depth-$depth.json --output-format=json --filename=/dev/$DEV --direct=1 --size=$SIZE --ioengine=libaio --iodepth=$depth --rw=randread --bs=4Ki --runtime=30 --numa_cpu_nodes=$DEV_NUMA_NODE --ramp_time=10 --time_based --percentile_list=50:95

# Namespace is still full so run overwrite benchs
$ sudo numactl -m $DEV_NUMA_NODE fio --name=$DEV_overwrite-seq_4Ki_queue-depth-$depth --output=$DEV_overwrite-seq_4Ki_queue-depth-$depth.json --output-format=json --filename=/dev/$DEV --direct=1 --size=$SIZE --ioengine=libaio --iodepth=$depth --rw=write --bs=4Ki --runtime=30 --numa_cpu_nodes=$DEV_NUMA_NODE --ramp_time=10 --time_based --percentile_list=50:95
$ sudo numactl -m $DEV_NUMA_NODE fio --name=$DEV_overwrite-rand_4Ki_queue-depth-$depth --output=$DEV_overwrite-rand_4Ki_queue-depth-$depth.json --output-format=json --filename=/dev/$DEV --direct=1 --size=$SIZE --ioengine=libaio --iodepth=$depth --rw=randwrite --bs=4Ki --runtime=30 --numa_cpu_nodes=$DEV_NUMA_NODE --ramp_time=10 --time_based --percentile_list=50:95
\end{lstlisting}

\begin{lstlisting}[language=bash,caption={Establishing the maximum achievable bandwidth of the ZNS convention namespace with I/O queue depth of 4 and increasing block size.},label={listing:conv_bandwidth}]
# We define several variables, replace these with values
    # DEV: device name (e.g., nvme2n1)
    # block_size: I/O size (from 4KiB to 128KiB in powers of 2)
    # DEV_NUMA_NODE: NUMA Node of the device
    # SIZE: device size (e.g., 100G)

$ sudo numactl -m $DEV_NUMA_NODE fio --name=$DEV_randwrite_$block_size_queue-depth-4 --output=$DEV_randwrite_$block_size_queue-depth-4.json --output-format=json --filename=/dev/$DEV --direct=1 --size=$SIZE --ioengine=libaio --iodepth=4 --rw=randwrite --bs=$block_size --runtime=30 --numa_cpu_nodes=$DEV_NUMA_NODE --ramp_time=10 --time_based --percentile_list=50:95

# Reset the namespace between write benchmarks (only on namespaces, not partitions)
$ sudo nvme format /dev/$DEV -f

# Run remaining benchmarks
$ sudo numactl -m $DEV_NUMA_NODE fio --name=$DEV_write_$block_size_queue-depth-4 --output=$DEV_write_$block_size_queue-depth-4.json --output-format=json --filename=/dev/$DEV --direct=1 --size=$SIZE --ioengine=libaio --iodepth=4 --rw=write --bs=$block_size --runtime=30 --numa_cpu_nodes=$DEV_NUMA_NODE --ramp_time=10 --time_based --percentile_list=50:95
$ sudo numactl -m $DEV_NUMA_NODE fio --name=$DEV_read_$block_size_queue-depth-4 --output=$DEV_read_$block_size_queue-depth-4.json --output-format=json --filename=/dev/$DEV --direct=1 --size=$SIZE --ioengine=libaio --iodepth=4 --rw=read --bs=$block_size --runtime=30 --numa_cpu_nodes=$DEV_NUMA_NODE --ramp_time=10 --time_based --percentile_list=50:95
$ sudo numactl -m $DEV_NUMA_NODE fio --name=$DEV_randread_$block_size_queue-depth-4 --output=$DEV_randread_$block_size_queue-depth-4.json --output-format=json --filename=/dev/$DEV --direct=1 --size=$SIZE --ioengine=libaio --iodepth=4 --rw=randread --bs=$block_size --runtime=30 --numa_cpu_nodes=$DEV_NUMA_NODE --ramp_time=10 --time_based --percentile_list=50:95

# Namespace is still full so run overwrite benchs
$ sudo numactl -m $DEV_NUMA_NODE fio --name=$DEV_overwrite-seq_$block_size_queue-depth-4 --output=$DEV_overwrite-seq_$block_size_queue-depth-4.json --output-format=json --filename=/dev/$DEV --direct=1 --size=$SIZE --ioengine=libaio --iodepth=4 --rw=write --bs=$block_size --runtime=30 --numa_cpu_nodes=$DEV_NUMA_NODE --ramp_time=10 --time_based --percentile_list=50:95
$ sudo numactl -m $DEV_NUMA_NODE fio --name=$DEV_overwrite-rand_$block_size_queue-depth-4 --output=$DEV_overwrite-rand_$block_size_queue-depth-4.json --output-format=json --filename=/dev/$DEV --direct=1 --size=$SIZE --ioengine=libaio --iodepth=4 --rw=randwrite --bs=$block_size --runtime=30 --numa_cpu_nodes=$DEV_NUMA_NODE --ramp_time=10 --time_based --percentile_list=50:95
\end{lstlisting}

\subsection{Zoned Device Performance}\label{appendix:Zoned_device_performance}
We run various workloads on the zoned namespace of the ZNS device, and compare the performance of the scheduler set to \textit{mq-deadline} and \textit{none}. Listing~\ref{Listing:ZNS_setup} showed how to change the scheduler for a namespace. For all benchmarks we utilize fio configured to use 4KiB I/Os. 

\noindent \textbf{Write performance:} The benchmarks are shown in Listing~\ref{listing:ZNS_baseline_performance}. The listing also shows the write benchmark that was used to produce Figure~\ref{fig:ZNS_concur_write_seq}. This benchmark writes a single 4KiB I/O to a zone and we increase the number of concurrent writes that are issued to the increasing number of active zones. Therefore, the benchmarks for the different schedulers have the exact same command, and only requires to change the scheduler in between iterations, when the number of concurrent jobs is increased. Note, we also use a 50 zone namespace and have a maximum of 14 active zones, which is why the \textit{offset\_increment} flag is set to 3z, such that each additional thread starts at an increasing offset of 3 zones and all 14 threads can still fit into the namespace when running concurrently. In addition, the write size of each thread is 3z. Also note that the benchmarks with I/O queue depth of 1 and increasing threads over active zones utilize the \textit{psync} ioengine, since I/Os are synchronous, as opposed to \textit{libaio} with asynchronous benchmarks where I/O queue depth is larger than 1. 

\begin{lstlisting}[language=bash,caption={Measuring write latency for the schedulers with increasing number of active zones and a single outstanding I/O per zone.},label={listing:ZNS_baseline_performance}]
# We define several variables, replace these with values
    # DEV: device name (e.g., nvme2n2)
    # DEV_NUMA_NODE: NUMA Node of the device
    # scheduler: current scheduler
    # jobs: [1-14] number of active zones

# Write benchmark, change the scheduler between iterations
$ sudo numactl -m $DEV_NUMA_NODE fio --name=$(echo "${DEV}_${scheduler}_4Ki_nummjobs-${jobs}") --output=$(echo "${DEV}_${scheduler}_4Ki_numjobs-${jobs}.json") --output-format=json --filename=/dev/$DEV --direct=1 --size=3z --offset_increment=3z --ioengine=psync --zonemode=zbd --rw=write --bs=4Ki --runtime=30 --numa_cpu_nodes=$DEV_NUMA_NODE --ramp_time=10 --group_reporting --numjobs=$jobs --percentile_list=50:95
\end{lstlisting}

The next write benchmark, as was shown in Figure~\ref{fig:ZNS_concur_write_seq_iodepth}, runs \textit{mq-deadline} with a single zone and an increasing I/O queue depth and \textit{none} runs just as before, with a single I/O per zone and an increasing number of active zones. Listing~\ref{listing:ZNS_concur_write_seq_iodepth} shows these specific commands. This benchmark defines the size as the total space of the device, since it runs a single thread.

\begin{lstlisting}[language=bash,caption={Measuring write latency for the schedulers with increasing number of active zones and a single outstanding I/O per zone.},label={listing:ZNS_concur_write_seq_iodepth}]
# In addition to the prior defined variables we also define
    # SIZE: device size (e.g., 100G)
    # depth: [1-14] I/O queue depth,
    #        and numjobs for none

# mq-deadline benchmark
$ sudo numactl -m $DEV_NUMA_NODE fio --name=$(echo "${DEV}_mq-deadline_${BS}_iodepth-${depth}") --output=$(echo "${DEV}_mq-deadline_${BS}_iodepth-${depth}.json") --output-format=json --filename=/dev/$DEV --direct=1 --size=$SIZE --ioengine=libaio --zonemode=zbd --rw=write --bs=4Ki --runtime=30 --numa_cpu_nodes=$DEV_NUMA_NODE --ramp_time=10 --iodepth=$depth --time_based --percentile_list=50:95

# none benchmark
$ sudo numactl -m $DEV_NUMA_NODE fio --name=$(echo "${DEV}_none_4Ki_nummjobs-${jobs}") --output=$(echo "${DEV}_none_4Ki_numjobs-${jobs}.json") --output-format=json --filename=/dev/$DEV --direct=1 --size=$SIZE --offset_increment=3z --ioengine=psync --zonemode=zbd --rw=write --bs=4Ki --runtime=30 --numa_cpu_nodes=$DEV_NUMA_NODE --ramp_time=10 --numjobs=$depth --time_based --group_reporting --percentile_list=50:95
\end{lstlisting}

\noindent \textbf{Read performance:} Read performance was measured with sequential reading of 4KiB on a single zone and an increasing I/O queue depth for both schedulers. Recall, that ZNS devices only have a sequential write constraint, and hence read requests can be issued with increasing I/O queue depth under any scheduler and do not have to be at the write pointer of a zone. Listing~\ref{listing:ZNS_concur_read_seq_iodepth} shows the command for this particular benchmark. The benchmark increases only the I/O queue depth from 1 to 14, and commands are the same under both scheduler configurations, however the correct scheduler has to be set before each iteration.

Next, we run the random read benchmark. The \textit{mq-deadline} configuration uses the exact same set up as for sequential reading, namely issuing 4KiB I/Os in a single zone with an increasing I/O queue depth. However, \textit{none} issues a single 4KiB I/O per zone, with an increasing number of concurrent threads. The command from Listing~\ref{listing:ZNS_concur_read_seq_iodepth} can be used for the \textit{mq-deadline} scheduler, and the \textit{none} scheduler can use the command in Listing~\ref{listing:ZNS_concur_write_seq_iodepth}, with the benchmark parameter \textit{--rw=randread}. Note, that similar to the prior read benchmark, the namesapce needs to be full in order to be able to read data.

\begin{lstlisting}[language=bash,caption={Measuring sequential read performance for 4KiB I/Os in a single zone with increasing I/O queue depth.},label={listing:ZNS_concur_read_seq_iodepth}]
# Defined variables
    # scheduler: current scheduler
    # depth: [1-14] I/O queue depth

# Change scheduler between iterations
$ sudo numactl -m $DEV_NUMA_NODE fio --name=$(echo "${DEV}_${scheduler}_4Ki_iodepth-${depth}") --output=$(echo "${DEV}_${scheduler}_4Ki_iodepth-${depth}.json") --output-format=json --filename=/dev/$DEV --direct=1 --size=SIZE --ioengine=libaio --zonemode=zbd --rw=read --bs=4Ki --runtime=30 --numa_cpu_nodes=$DEV_NUMA_NODE --ramp_time=10 --iodepth=$depth --time_based --percentile_list=50:95
\end{lstlisting}

\end{document}